\documentclass{sig-alternate}
\usepackage[numbers,sort&compress]{natbib}
\usepackage[usenames,dvipsnames]{color}
\usepackage{amsmath}
\usepackage{times,xspace,subfigure,url,graphicx,grffile} 
\usepackage{balance}
\usepackage{array}
\usepackage{enumerate}
\usepackage{multirow}
\usepackage{rotating}

\newcommand\Focus{Focus\xspace}
\newcommand\focus{focus\xspace}
 
\newcommand\focused{focused\xspace} 
\newcommand\FOCUSED{FOCUSED\xspace} 

\begin{document}
\conferenceinfo{WSDM'13,} {February 4--8, 2013, Rome, Italy.}
\CopyrightYear{2013}
\crdata{978-1-4503-1869-3/13/02} 
\widowpenalty=10000\clubpenalty=1000

\title{Characterizing and Curating Conversation Threads: Expansion,
  \Focus, Volume, Re-entry
}

\numberofauthors{1} 
\newcommand{\authspace}{\hspace*{.2in}}
\author{
\begin{tabular}{cccc}
Lars Backstrom &  Jon Kleinberg &  Lillian Lee  & \mbox{Cristian Danescu-Niculescu-Mizil}\\
 \affaddr{Facebook} &  \affaddr{Cornell University} & \affaddr{Cornell University} & \affaddr{Stanford/Max Planck Institute SWS} \\
 \affaddr{lars@fb.com} &   \affaddr{kleinber@cs.cornell.edu} &  \affaddr{llee@cs.cornell.edu} &   \affaddr{cristiand@cs.stanford.edu}
\end{tabular}
}

\maketitle
\begin{abstract}

Discussion threads form a central part of the experience on many Web
sites, including social networking sites such as Facebook and Google
Plus
 and knowledge creation sites such as Wikipedia. To help users
manage the challenge of allocating their attention among the
discussions that are relevant to them, there has been a growing need for
the algorithmic curation of on-line conversations --- the development
of automated methods to select a subset of discussions to present to a
user.

Here we 
consider
two key sub-problems inherent in 
conversational curation: length prediction --- predicting the number
of comments a discussion thread will receive --- and 
the novel task of 
re-entry
prediction --- predicting whether a user who has participated in a
thread will later contribute another comment to it.  The first of
these 
sub-problems
arises in estimating how interesting a thread is,
in the sense of generating a lot of conversation;
the second can help determine
whether users should be kept notified of the progress of a thread
to which they have already contributed.
We develop and evaluate a range of
approaches
for these tasks, based on an analysis of the network structure
and arrival pattern among the participants, as well as a novel
dichotomy in the structure of long threads. We find that for both
tasks, learning-based approaches using these sources of information
yield improvements for all the performance metrics we used.

\end{abstract}

\smallskip

\noindent {\bf Categories and Subject Descriptors}:
H.2.8: Data Mining 

\smallskip

\noindent {\bf General Terms:} Measurement; Experimentation; Theory

\smallskip


\def\unif{{\cal U}}
\def\act{{\cal A}}
\def\pop{{\cal P}}
\newcommand{\omt}[1]{}
\newcommand{\xhdr}[1]{\paragraph*{\bf #1}}
\newcommand{\yhdr}[1]{\subsection*{\bf #1}}
\def\T{{\cal T}}
\def\L{{\cal L}}
\renewcommand{\^}[1]{^{(#1)}}

\section{Introduction}
\label{sec:intro}

Many Web sites are organized around a continuously
evolving set of discussion threads.
This style of interaction is a key
component of on-line groups and message boards,
social networking sites such as Facebook and Google Plus, 
and the workflow of collaborative projects such as Wikipedia
and open-source development.
In all these cases, a user must continuously
decide how to allocate his or her attention 
to a range of relevant discussions,
and this can be a challenging task when content arrives at a rapid rate.

A growing number of sites are helping users address this challenge
through the algorithmic {\em curation} of discussion threads,
automatically selecting which threads to bring to a user's
attention at any given point in time.
A canonical example is Facebook's News Feed --- for users with
a sufficient number of active friends on the site, an unfiltered
stream of all stories generated by friends 
is generally much less valuable to the user than a ranked and filtered 
version of the stream that attempts to highlight the stories 
estimated to be most engaging to the user.

The problem of curating discussion threads is thus a wide-ranging one
in the context of applications, 
but it is one which 
for the most part
has not been systematized in prior
research. 
Our goal in this paper is to facilitate
such a systematization, by identifying 
and formalizing two important sub-problems in conversational curation, 
and then developing and evaluating techniques to address them.
For our evaluation, we use discussion threads from two sites where
such threads form a core part of the experience: discussions
among users on Facebook and discussions among editors on Wikipedia.
As on many other sites, threads on Facebook and Wikipedia
can be conceptualized as 
an initial {\em post} and a subsequent sequence of
{\em comments}; we will use this terminology in what follows.

\xhdr{The present work: Two problems in conversational curation}
We now describe the two problems that we study, together with their
motivation as components of conversational curation.
\begin{enumerate}
\item 
{\em Length prediction}: given the
initial portion of a thread (a post and the 
first few comments following it),
how well can we predict the eventual length of the thread?
We use this length prediction problem as a concretely formulated
proxy for the general issue of estimating the level of interest a
thread will generate, based on observation of its early stages.
\item
{\em Re-entry prediction}: given the
initial portion of a thread and the identity of one of the commenters,
how well can we predict whether this commenter will contribute another
comment later in the thread? 
This is a key issue in determining whether to keep a user notified
of the progress on a thread once he or she has contributed to it ---
some threads have the structure of a conversation where users are
motivated to return repeatedly, while others involve each user
contributing once (for example, to offer congratulations or 
condolences) but then not returning.  
\end{enumerate}
Taken together, these two problems cover a set of central issues
in conversational curation: identifying threads that will generate
sustained interest,
so as to be able to highlight them to users,
and
recognizing whether a thread is something that a contributing
user will want to continue to follow as it evolves.

We develop techniques for these problems by first analyzing the
structure of threads, and then formulating a set of properties that
we in turn use for the prediction tasks.

\newcommand{\secbold}[1]{{\bf #1}}
We begin by investigating the following issue, 
on data described in \secbold{\S\ref{sec:data}}.
Intuitively, one feels from experience that there are two distinct
types of long threads: those that become long because a small group
of people engage in an extensive conversation via the comments,
and those that become long because many users each contribute a
single comment.  
A canonical example of the latter would begin
with a post in which a user announces a major life event, and then
many friends contribute congratulations in the comments
as in a wedding guestbook.
We 
refer to the first type of thread
as {\em focused}, and the second type as {\em expansionary}.

But is this notion of two types simply one's perception of two
extremes of a broad distribution, or is there quantitative evidence for it?
We find  
(\secbold{\S\ref{sec:dichotomy}}) in fact that threads genuinely exhibit this two-type
effect: for long threads, the distribution of the number of distinct
commenters is bimodal, with threads either dominated by
a very small number of distinct users, or by a sequence of commenters who
generally do not return to the thread after commenting once.
In addition to providing what is, to our knowledge,
the first evidence for this basic dichotomy, 
this finding helps reinforce the importance of our second problem ---
re-entry prediction --- by establishing that active discussion 
threads can vary considerably in the extent to which participants
are interested in returning after their initial contribution.

In order to build a framework for approaching our two basic problems, we 
begin by studying (\secbold{\S\ref{sec:participants}}) 
a range of related thread properties.
One of the most useful of these is the thread's {\em arrival pattern} ---
the ordering by which new entrants into the thread are interleaved
with returning participants.  Formalizing this notion allows us
to work with relaxed versions of the two extremes of
focused and expansionary threads discussed above,
and to explore the region that interpolates between them.
We also study network and temporal structure: whether the first
few commenters are linked
within a
social network, and how quickly 
after the post 
do
they arrive in real-time; both convey information about the
future trajectory of the thread.

We incorporate these properties into a machine learning approach 
for predicting length (\secbold{\S\ref{sec:predict-length}}) and re-entry (\secbold{\S\ref{sec:predict-reentry}}).  Evaluating the prediction
performance enables us to identify the features that are most
effective for our two problems.
At a high level, we find that the structure of the arrival pattern
is the most useful for re-entry prediction, while temporal properties
together with the arrival pattern give the strongest performance
for length prediction.

Next, in \secbold{\S\ref{sec:model}},
we explore a 
probabilistic model
of participant re-entry
related to the dichotomy 
between focused threads 
and expansionary ones.
Clearly some styles of post tend to lead to one type
of thread or the other,
but for other kinds of posts,
one sees both types of threads emerge;
for example, 
the same shared link to a news story can generate a focused
thread when it is shared among one set of users and
an expansionary thread when it is shared among a different set.
It is therefore natural to ask whether a type of symmetry-breaking can
arise directly from the dynamics of a discussion itself --- that is,
whether there is a simple probabilistic generative model capable of producing 
both focused and expansionary threads over different realizations of its
random trajectory.  We show how to construct such a model from 
plausible assumptions about turn-taking and new entrants in discussion
threads; the model exposes interesting connections between discussion
threads and nonlinear urn processes.

In \secbold{\S\ref{sec:relwork}}, we review related work on the dynamics of
on-line discussions. For now, we note that the general issue of thread 
length has been studied, using different techniques, in contexts
distinct from ours --- primarily for comments on blog and news sites,
where essentially all threads are expansionary, with many participants
who typically contribute only once or very few times each
\citep{Tsagkias:2009:PVC:1645953.1646225,Guerini:ProceedingsOfIcwsm:2011,Yano:ProcOfIcwsm:2010,Wang:ProceedingsOfKdd:2012}.
In contrast, our approach incorporating the notion that there can
be multiple structurally distinct types of long threads is suited
to settings where the participants maintain long-running relationships
with one another.
These structural distinctions also provide a core part of the motivation
for re-entry prediction, which is a key issue for organizing
conversations in these settings; the problem of re-entry prediction
has not, to our knowledge, been formulated or studied previously.

\def\W{{\cal W}}

\section{Data and Basic Definitions}
\label{sec:data}

We use data from Facebook and Wikipedia to construct three distinct
populations of users whose discussion threads we study.  We choose
Facebook as perhaps the most well-known example of a
post-plus-comments interface for socially-oriented conversations. 
Conversations among Wikipedia editors form a
contrasting case that has also received research attention \citep{Laniado+al:2011a,Gomez:ProceedingsOfTheAcmConferenceOnHypertext:2011,Danescu-Niculescu-Mizil:2012:EPL:2187836.2187931}: the discussions are task-oriented, as opposed
to socially-oriented, and there is no
formal structure imposed on conversations by the interface; 
nonetheless, they
can still be naturally treated as instances of comment threads.

For completeness, we briefly describe the structure of these discussion
threads at a general level.  
On Facebook, we study 
instances in which a user posts
a status update,
and then other users with permission
to comment on the status update contribute comments to it.
On Wikipedia, editors interact on 
{\em talk-pages}
to discuss issues concerning articles, projects or Wikipedia
policies.
Each editor has the option of hosting a talk-page, and most active
users do.

On both Facebook and Wikipedia, we will refer to the status update or
initiating text as the {\em post}; the sequence of comments
that follows the post will be called the {\em comment thread}, and 
the post together with all the comments will be called the 
{\em full thread}.  The poster together with the commenters in 
a full thread will be called the full thread's {\em participants}.
The number of 
items in the thread (including the post in the case
of a full thread, 
but not in a comment thread) will be called its {\em length} or its {\em volume};
we use these two terms synonymously.

From Facebook, we first selected 100,000 users uniformly at
random from the population of US Facebook users.
We will refer to this set $\unif$ in our analysis as the {\em uniform} 
Facebook population.
Also, out of all US Facebook users who posted beween 200 and 300
status updates over 
an 80-day period,
we randomly
selected 100,000 of these heavily engaged users.
We will refer to this set $\act$ as the 
{\em high-activity} Facebook population.
For both $\unif$ and $\act$, we study the comment threads associated 
with all their posts during 
the same 80-day period.
All Facebook data was used anonymously,
and all analysis was done in aggregate.

Our Wikipedia data is derived from the corpus of
\citet{Danescu-Niculescu-Mizil:2012:EPL:2187836.2187931}.
We used 118,447 conversation threads of length at least 1 (to discard posts made by automated bots, which never attract responses)
which took place asynchronously
on the talk-pages hosted by 6,555 highly active editors;
posts  average  2.12 comments.  We also use the content of the
talk-pages to
 assess the existence of an interactional link between a
given pair of Wikipedia editors: we say that two editors are linked if
at least one of them added a post or comment on the other's editor
talk page. Our Wikipedia data will be available at  \\
{\small \verb#http://www.mpi-sws.org/~cristian/Echoes_of_power.html#}.

\renewcommand{\yhdr}[1]{\xhdr{#1}}

\section{\FOCUSED vs. Expansionary Threads}
\label{sec:dichotomy}

If a post leads to a long comment thread, then 
it is one that attracts a great deal of attention
and so is likely of interest; 
thus, the thread-length
prediction problem is crucial to the curating of conversations.  
In thinking about how to bring 
long threads to users' attention, though, a natural question
is whether there are sub-classes of such conversations
that should be treated differently.

This question leads us to conjecture that there is a dichotomy between
{\em expansionary} high-activity threads, created by the one-time
actions of many different ``drive-by'' commenters,  versus {\em \focused}
high-activity threads, reflecting  a high-level of repeated
engagement among relatively few people.  In this section, we provide supporting
evidence for this conjecture and discuss its consequences.

\xhdr{Distinct Participants: Two Local Maxima} To investigate the
validity of our conjecture, we consider how
the number of distinct participants
in a thread is distributed.  
To do so, we must account for two issues.  First,  we do not
want the idiosyncratic actions of any one high-volume user to dominate
the quantities involved, so we work with a macro-averaged
function.\footnote{The results turn out to be
similar for the micro-averaged analog.} Second,  the possible number of distinct
participants in a thread depends on the thread's length, and so we
need to parametrize by it.

Thus, formally, 
for a population of users $\pop$, let 
$\pop_k$ be the set of users who authored at least one
post having comment thread length at least $k$.
For each user $u \in \pop_k$, we take all full threads associated with
a post by $u$ that produced at least $k$ comments,
and we truncate each of these threads to the prefix consisting
of just the post and the first $k$ comments.
Let $\delta_u(k)$ be the average number of distinct participants
in all these prefixes of full threads initiated by $u$.
(For a given
such prefix, the number
can range from 1 --- the original poster contributed all of the
comments as well --- to $k+1$ --- all commenters are distinct, and the
original poster didn't comment.)
We then define 
$\Delta_k^*(d)$ to be the fraction of users $u \in \pop_k$
for whom $\lfloor \delta_u(k) \rfloor = d$.
Note that $\Delta_k^*$ is a density function. 
In what follows, for brevity we will sometimes refer to it
simply as an average or an expectation, with the 
understanding that this refers in fact to a macro-averaged quantity.

\begin{figure}[t]
  \begin{center}
\hspace*{-.1in}\begin{tabular}{|l@{}m{.3in}|} \hline
\vtop{
  \vskip-8ex
  \hbox{
\includegraphics*[width=0.45\textwidth,viewport=30 30 645 420]{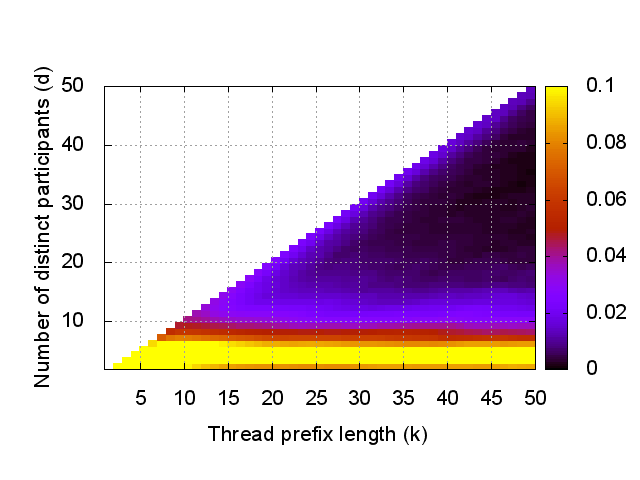}
}
}&\vspace{60pt}\hspace*{-.05in}{$\Delta_k^*(d)$}\\ \hline
\end{tabular}
    \caption{
Heat map (best viewed in color) of the density functions on distinct
commenters,  uniform Facebook population.  For each 
thread prefix
length $k$,
there is a peak 
in density (lighter color)
at a small number and a second peak approximately at 
the maximum number of distinct
participants $d$.
    \label{fig:hm-random}
    }
  \end{center}
  \vspace*{-0.02\textheight}
\end{figure}

In these terms, our conjecture can be 
expressed as follows: 
for threads of sufficient length $k$, the density function
$\Delta_k^*(d)$ should have (at least) two local maxima: one at a small
value of $d$, i.e., $d \ll k$,
and one at a large value of $d$, i.e., $d \approx k+1$.

In Figure \ref{fig:hm-random}, we show the family of density functions
$\Delta^*_k$ for $k \in 1, 2, ..., 50$ on our uniform 
Facebook population $\unif$.
The densities are drawn as a heat map, with column $k$ representing
the density function $\Delta_k^*$.
We see that as $k$ increases, $\Delta^*_k(d)$ is
first
 maximized at $d = 4$,
reflecting the dominant role of the \focused effect;
but then,
a second local maximum emerges at a value of 
$d$ very close to $k$.
For the population  $\act$ of high-activity Facebook users, we see
essentially the same effect, including the two local maxima at $d = 4$
and $d$ close to $k$ (figure omitted for space).

Although the smaller data volume makes it more difficult to
discern the effect on Wikipedia, when we group together
the possible values of $d$ into 
contiguous intervals,
we find significant
evidence of two local maxima there too.
To quantify the effect on Wikipedia, 
we compare quantiles of $\Delta_k^*$, defining 
$\displaystyle{ f_k(p,q) = \sum_{pk \leq d \leq qk} \Delta_k^*(d) }$.
We find that as $k$ increases
(in particular, considering $k \geq 15$),
we have $f_k(0,\frac14) > f_k(\frac14,\frac12)$ and 
$f_k(\frac34,1) > f_k(\frac12,\frac34)$.
This inequality is consistent with Figure \ref{fig:hm-random},
where the density function is larger at the two extremes than in
comparably-sized intervals in between.

\xhdr{Consequences: Predict Both Length and Re-entry} 
As argued earlier, conversational-curation systems should contain a
thread-length prediction component.  But our new observation about
the distinction between expansionary and \focused threads 
shows that long threads can differ significantly in the extent
to which a commenter will want to return to contribute a second time.
This heterogeneity in long threads motivates the formulation
of our second task, {\em re-entry prediction}:
determining whether a given participant in the thread is
likely to contribute again.
To our knowledge, re-entry in on-line conversations is a problem
that has not been previously formalized or studied.

Length and re-entry are important, and distinct, issues in 
the task of conversational curation.
Length prediction, since it provides information about the amount of attention 
a thread is likely to receive, helps in assessing whether a user
should be made aware of the thread at all.
Re-entry prediction, on the other hand, provides information 
about how to keep a user informed of the evolution of the thread
once he or she has already contributed to it: 
a high re-entry probability indicates that the user may well want
to know about subsequent comments, so that he or she can contribute
in response to them.

Predicting a particular user's re-entry is different
from predicting whether the thread itself will be focused or expansionary.
While very few users re-enter an expansionary thread by definition,
it is easily possible for a user $u$ to contribute to a thread
that later becomes dominated by a back-and-forth discussion among
a small set of other participants; in this case, the thread is focused,
but user $u$'s re-entry probability might be low.
Predicting re-entry provides a concrete recommendation with
respect to a given user, in a way that predicting whether
a thread will be focused or expansionary does not.

We note that re-entry prediction is focused on a user's {\em production}
of comments --- specifically, whether the user will write another
comment in the future.
An interesting open question is to consider the analogous prediction
task for a user's {\em consumption} of comments.
In particular, a user might be interested in continuing to read
comments on a thread as they come in, despite having no intention
of contributing again.
(Consider a string of congratulatory messages on a life event that 
include interesting side information, such as personal reminiscences
or clever quips.)

\newcommand{\userone}{\textit{user1}\xspace}
\newcommand{\nn}{\#} 
\section{Early participants: social, sequential, and temporal structures}
\label{sec:participants}

In this section, we show how properties of the initial participants
in a thread can provide information about the thread's later dynamics,
thus laying the groundwork for the features in our subsequent
prediction experiments.
First (\S\ref{sec:links}), 
we show that
the presence or
absence of social links among the initial participants 
in a thread turns out to 
provide useful information, though 
in interestingly different ways for different settings. 
Second (\S\ref{sec:patterns}), 
inspired by our expansionary vs. \focused
analysis in \S\ref{sec:dichotomy},
which
 introduces the importance
of re-entry,  we develop a novel representation
for the sequence of participant contributions. 
Third (\S\ref{sec:timing}), 
we demonstrate that
how fast
the initial commenters arrive provides important information about
the eventual number of comments, though its connection
with re-entry probability is less clear.

\subsection{Links Among the Initial Participants}
\label{sec:links}

We 
first consider
comment threads in 
our Facebook High-Activity population (outcomes are analogous for the Uniform set), 
focusing on 
threads 
with at least two comments and
where the first
two commenters are distinct from each other and from the post's author.
The tension between the \focused and the expansionary effects 
has a natural reflection in the relationship between these first
two commenters.  If they are friends, then 
interest in the post might be limited to a particular portion of
the poster's social neighborhood. That is, interest
in such a post could have limited {\em reach},  which could restrict thread
length.  At the same time, though, there might also be
increased potential for an extended conversation to ensue as friends
interact, which would lead to a longer thread.

The top-left plot in 
Figure \ref{fig:first-connect} shows that in fact, these Facebook threads are significantly longer
when the first two commenters are friends.

We can further validate this hypothesized effect of
conversational interaction
by examining a related mechanism in which
the role of interaction is much more limited.
The Facebook ``like'' feature is very useful for this purpose.
Users can respond to a post not just by commenting on it but also by 
clicking the {\em like} (thumbs-up) button, 
which provides a one-bit endorsement 
of the content.
Thus, likes are a light-weight 
communication alternative to comments, and we can consider ``like
threads'' --- the sequence of likes arriving on a post --- as the
corresponding analog of comment threads.  But there is a crucial difference:
in like threads, there is no analog to the back-and-forth interaction that
characterizes conversational interaction.  

When the first two likers in a like thread are distinct, 
how is the eventual length of the like thread
affected by whether these two users are friends?
The top-middle plot of 
Figure \ref{fig:first-connect}(b) shows that, 
in the absence of repeated interactions to offset its consequences,
a limited-reach effect is clear: the like thread is shorter
when 
 the first two likers are linked.\footnote{We note that measuring the
number of {\em distinct} commenters shows the same limited-reach effect:
although the comment thread is longer when the first two commenters
are linked, the total number of distinct commenters is smaller.}

Applying the same analysis to Facebook thread prefixes of length
$k=3, 4,$ and $5$ yields very similar results.
For space reasons, we only depict the case $k=3$ (bottom-left and bottom-middle plots in
Figure \ref{fig:first-connect}), but the results are that expected
length of comment threads 
continues to increase almost perfectly monotonically in 
the number of edges among the first $k$ commenters when they are all
distinct.  Also, we find completely analogous results for re-entry in 
Facebook threads: the re-entry of the first participant increases
strongly with the number of edges among the first $k$ commenters
when they are all distinct.

What about Wikipedia comment threads?  As depicted in the rightmost
column of Figure 
\ref{fig:first-connect}, in this domain, it is 
{\em
not} the case that more connections between the first participants leads to
longer threads, although the available data here is quite sparse.  We
conjecture that the root cause for the striking contrast 
to Facebook
may be the
task-oriented nature of the setting, in which conversations may be less
discursive, and editors who have interacted in the past may
be more conversationally efficient in reaching a conclusion.

It is interesting to note that earlier work of Ugander et al. considered 
the level of connectedness among the set of users who appear in an invitation to join Facebook \cite{Ugander:ProceedingsOfTheNationalAcademyOfSciences:2012}; invitations that displayed users who were not linked to each other had higher overall conversion rates than invitations that displayed linked users. While invitations and comment threads are clearly different in nature, they both involve opportunities to engage a user in the activity of the site; whether there is a deeper relationship between the connectedness of commenters here and the connectedness of inviters in that setting is an interesting open question.

{
\newcommand{\figwidth}{1in}
\begin{figure}
\hspace*{-.0in}\begin{tabular}{c|c|c|c}
  & (a) FB comments & (b) FB likes & (c) WK comments \\ \hline
{\rotatebox{90}{~ $k = 2$}} &
\includegraphics[width=\figwidth]{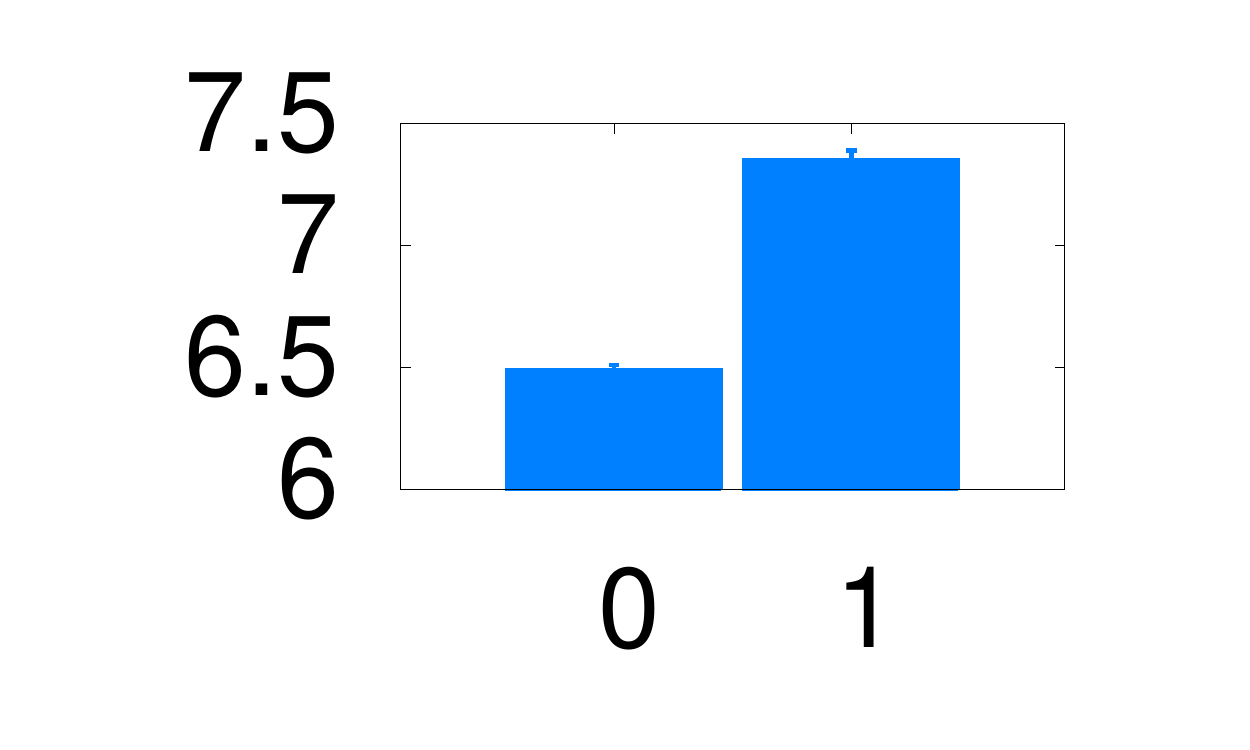} &
  \includegraphics[width=\figwidth]{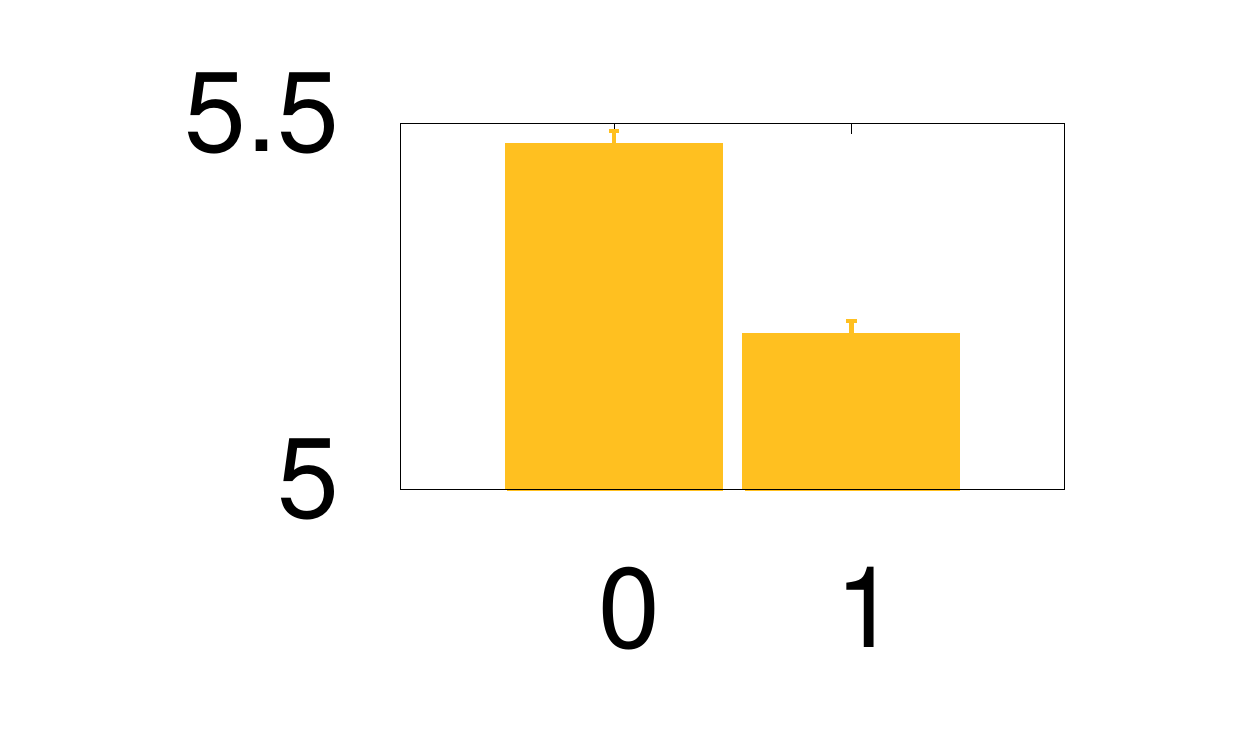}  &
  \includegraphics[width=\figwidth]{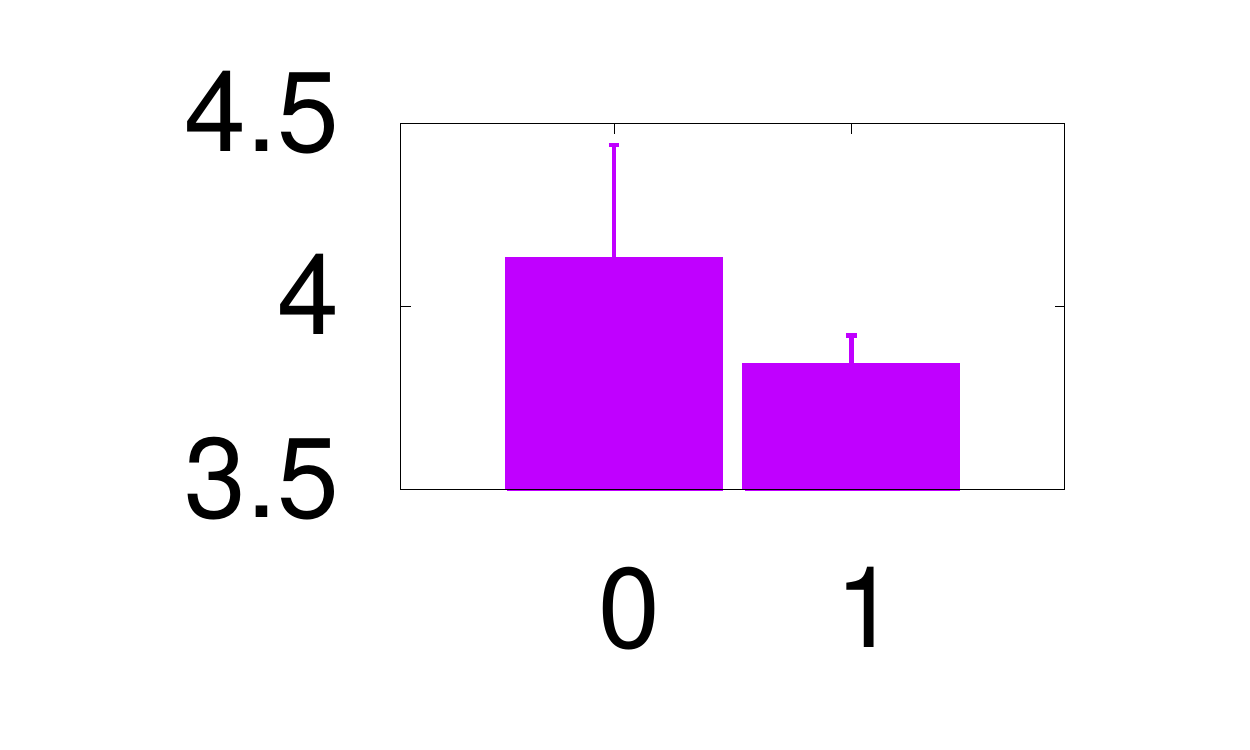} \\ 
\hline {\rotatebox{90}{~ $k = 3$}} &  \includegraphics[width=\figwidth]{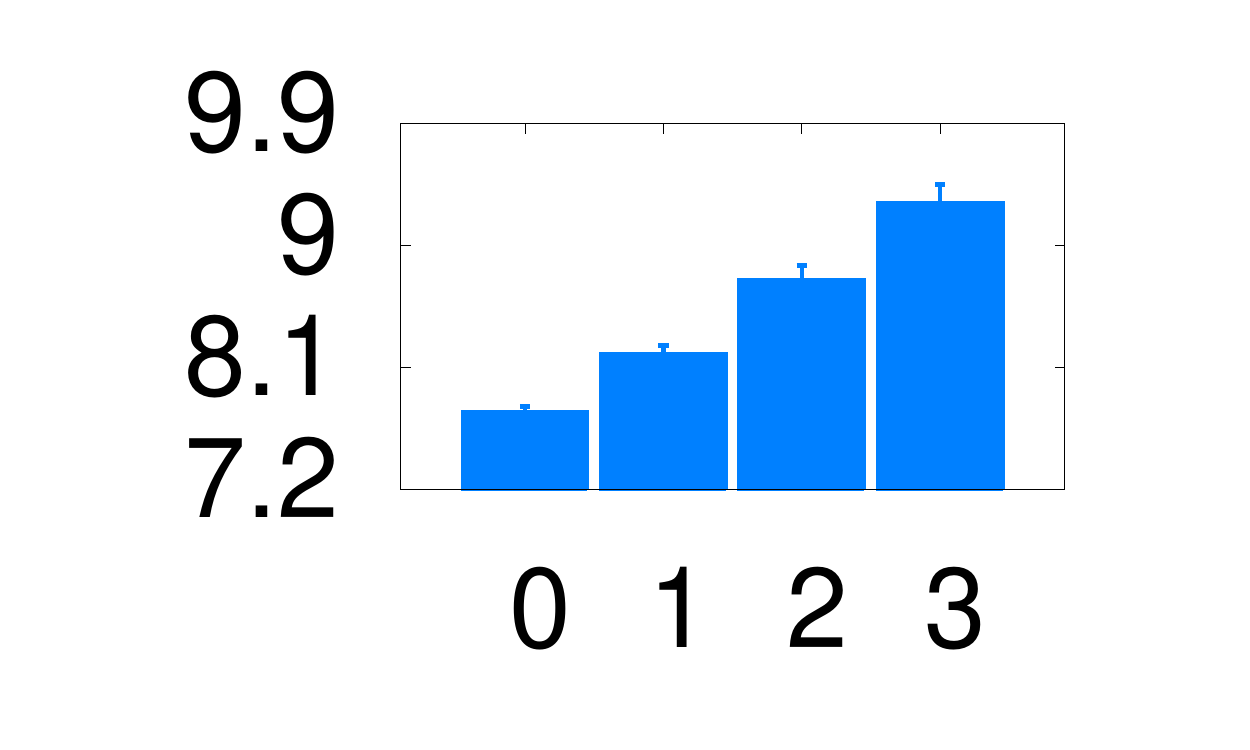} &
  \includegraphics[width=\figwidth]{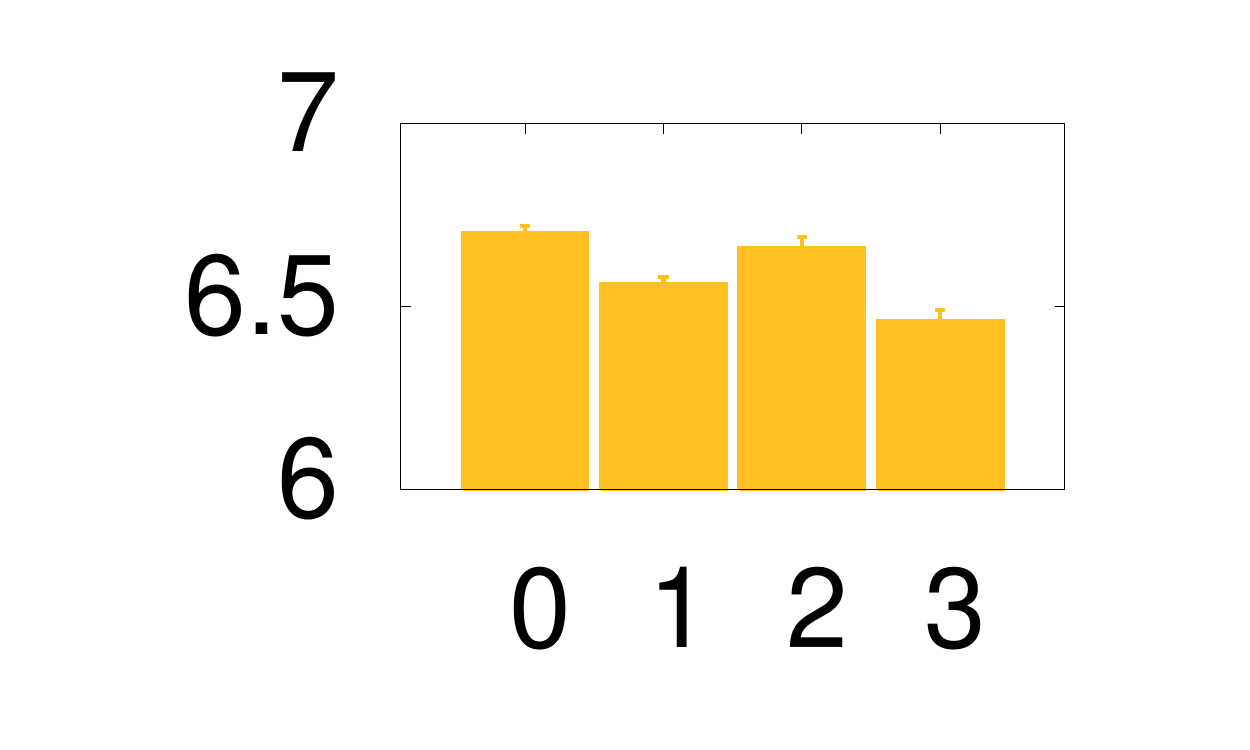}  &
  \includegraphics[width=\figwidth]{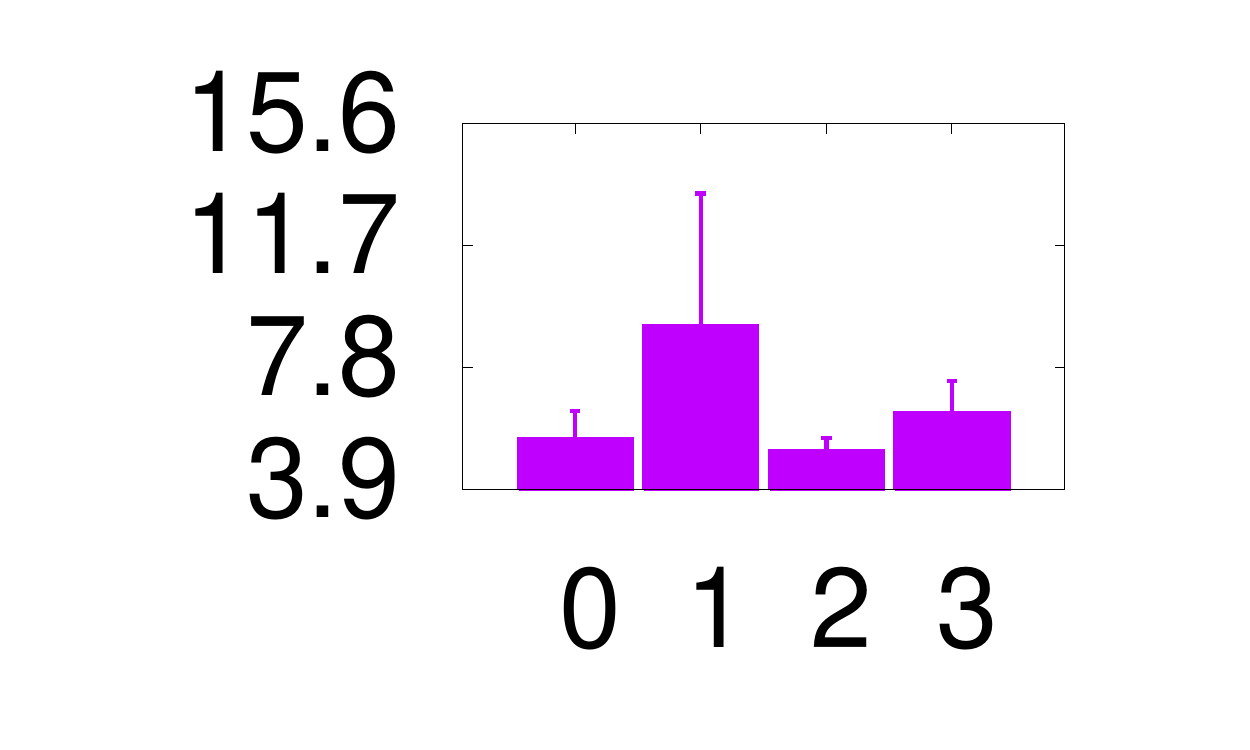}  
\end{tabular}
\caption{\label{fig:first-connect} Thread length vs. number of
  connections between the first $k$ commenters when they are distinct.
More-connected commenters make  for longer Facebook (FB) comment threads but
  shorter ``like threads'' and Wikipedia (WK)
  discussion threads.  (Y-axes not aligned to heighten trend visibility.)
\vspace*{-0.02\textheight}
}
\end{figure}
}

\subsection{Arrival Patterns}
\label{sec:patterns}

\newcommand{\relfreq}{\% of occ.}
\newcommand{\arrpattern}{Pattern}
\newcommand{\reenter}{1 re-enters}
\newcommand{\marg}{MargErr}

\newif{\ifpattern}
\patterntrue

\newcommand{\alldist}[1]{{\bf #1}}
\begin{table*}[t]
\begin{center}
{\small 
\ifpattern
\begin{tabular}{ccc}
\else
\begin{tabular}{cc||c}
\fi
\begin{tabular}{|l| r@{\%}|r|} \hline
\multicolumn{3}{|c|}{Facebook High-Activity, Length-5 arrival patterns } \\ \hline
\ifpattern
pattern & \multicolumn{1}{c|}{\reenter}&\relfreq \\   \hline \hline
\else
pattern bins & \multicolumn{1}{c|}{\reenter}&\relfreq \\   \hline \hline
\fi
\ifpattern
1,0,1,0,1 & 55.2 & 19.2 \\ \hline
1,0,1,0,0 & 47.5 & 2.8 \\ \hline
1,0,1,0,2 & 26.7 & 4.9 \\ \hline
1,0,1,2,0 & 26.1 & 4.1 \\ \hline
1,0,2,0,2 & 16.5 & 4.6 \\ \hline
1,2,0,2,0 & 14.6 & 3.1 \\ \hline
1,0,2,3,0 & 12.5 & 1.9 \\ \hline
1,2,0,3,0 & 11.5 & 2.0 \\ \hline
1,0,2,0,3 & 10.9 & 2.6 \\ \hline
1,2,3,4,5 & 5.6 & 3.6 \\ \hline
\else
 \nn0:2 \nn1:3& 55.2 & 19.2 \\ \hline
 \nn0:3 \nn1:2 & 47.5 & 2.8 \\ \hline
\nn0:2 \nn1:2 \nn2:1 & 26.4 & 9.0 \\ \hline
\nn0:2 \nn1:1 \nn2:2 & 15.7 & 7.7 \\ \hline
\nn:0:2 \nn1:1 \nn2:1 \nn3:1 & 11.6 & 6.5 \\ \hline
 \alldist{1,2,3,4,5} & \alldist{5.6} & {3.6} \\ 
\fi
\hline
  \multicolumn{2}{c}{}& \multicolumn{1}{|r}{sum: 48.8} 
\ifpattern \else
\\ \multicolumn{3}{c}{}\\ \multicolumn{3}{c}{}\\ \multicolumn{3}{c}{}\\ \multicolumn{3}{c}{}\\ 
\fi
\end{tabular}
&
\begin{tabular}{|l| r@{\%} |r|} \hline
\multicolumn{3}{|c|}{Wikipedia, Length-5 arrival patterns} \\ \hline
\ifpattern
pattern &\multicolumn{1}{c|}{\reenter} & \relfreq\\ \hline \hline
\else
pattern bins  &\multicolumn{1}{c|}{\reenter} & \relfreq\\ \hline \hline
\fi
\ifpattern
1,0,1,1,0 & 60.2 & 2.4 \\ \hline
1,1,0,1,0 & 58.6 & 2.8 \\ \hline
1,0,1,0,0 & 55.3 & 4.8 \\ \hline
1,0,0,1,0 & 52.2 & 5.5 \\ \hline
1,1,1,1,1 & 47.5 & 2.3 \\ \hline
1,0,1,2,1 & 46.0 & 3.2 \\ \hline
1,0,1,0,2 & 45.8 & 2.7 \\ \hline
1,2,1,2,1 & 41.1 & 5.1 \\ \hline
1,0,1,0,1 & 38.8 & 27.0 \\ \hline
1,2,3,4,5 & 7.2 & 1.7 \\ \hline
\else
\nn0:3 \nn1:2 & 53.6 & 10.3 \\ \hline
\nn0:2 \nn1:3 & 49.7 & 32.2 \\ \hline
 \nn1:5 & 47.5 & 2.3 \\ \hline
 \nn0:1 \nn1:3 \nn2:1 & 46.0 & 3.2 \\ \hline
 \nn0:2 \nn1:2 \nn2:1 & 45.8 & 2.7 \\ \hline
 \nn1:3 \nn2:2 & 41.1 & 5.1 \\ \hline
 \alldist{1,2,3,4,5} & \alldist{7.2} & {1.7} \\ \hline
\fi
 \multicolumn{2}{c|}{} & \multicolumn{1}{r}{sum: 57.5} \\ 
\ifpattern \else
 \multicolumn{3}{c}{}\\  \multicolumn{3}{c}{} \\   \multicolumn{3}{c}{}
\fi
\end{tabular}
&
{
\begin{tabular}{|l|r@{\%}|r|} \hline
\multicolumn{3}{|c|}{Facebook High-Activity, length-9 arrival patterns} \\ \hline
pattern bins& \multicolumn{1}{c|}{\reenter} & \relfreq\\ \hline \hline
\nn0:3, \nn1:6 & 67.7 & 1.7 \\ \hline
\nn0:4,  \nn1:5 & 66.9 & 12.1 \\ \hline
\nn0:5, \nn1:4 & 65.5 & 4.0 \\ \hline
\nn0:3, \nn1:4,  \nn2:2 & 56.8 & 2.1 \\ \hline
\nn0:3, \nn1:3,  \nn2:3 & 50.9 & 1.7 \\ \hline
\nn0:4, \nn1:4,  \nn2:1 & 47.6 & 5.2 \\ \hline
\nn0:4, \nn1:3,  \nn2:2 & 38.5 & 3.5 \\ \hline
\nn0:4, \nn1:3,  \nn2:1,  \nn3:1 & 28.2 & 2.0 \\ \hline
\nn0:4, \nn1:2,  \nn2:3 & 22.3 & 2.8 \\ \hline
\nn0:4, \nn1:1,  \nn2:4 & 9.6 & 2.7 \\ \hline
\multicolumn{2}{c|}{} & \multicolumn{1}{c}{sum: 37.8} \\ 
\end{tabular}
 } 
\end{tabular}
} 
\end{center}
\ifpattern
\vspace*{-0.02\textheight}
\fi
\caption{\label{tab:wiki_code_5} 
\ifpattern
{\em Left} and {\em middle}: the most common 
length-5 arrival patterns on Facebook, accounting for 48.6\% of the occurrences of all possible such arrival patterns, and on Wikipedia, 
accounting for 57.5\% of all occurrences of all possible such patterns.
The patterns are sorted by
the percentage of corresponding threads in which the user
with ID code 1 returns to the thread to comment again.  ``\relfreq'':
percentage of threads of length $\geq 5$ prefixed by that pattern.   
{\em Right}: The same for the most
common length-9 arrival patterns,
except that patterns have been
binned by counts of ID codes since there are many possible length-9
patterns. For example,  "\nn0:3, \nn1:6" $=$  the set of length-9
patterns where ID code 0 occurs 3 times and ID code  1 occurs 6 times, 
in any order.
(Some populations omitted for brevity or due to data sparseness.)
\else
{\em Left and middle}: The ten most common length-5 arrival patterns (prefixes)
in the Facebook active users and the Wikipedia populations, binned by counts of ID codes for brevity:  for example,  in Wikipedia, both 1, 0, 1, 0, 1 and 1, 0, 0, 1, 1 are in bin \nn0:2 \nn1:3, since in both, ID code 0 occurs twice and 1 three times.  
Bold: the canonical expansionary arrival pattern.
The bins are sorted by
the percentage of corresponding threads in which the user
with ID code 1 returns to the thread to comment again.  ``\relfreq'':
percentage of threads of length $\geq 5$ prefixed by that pattern (bin).   {\em Right}: The same for the most
common length-9 arrival patterns.
(Some populations omitted for brevity or due to data sparseness.)
\fi
}
\end{table*}

Having looked at the {\em number} of distinct commenters, and at
the {\em graph structure} on the first few participants in the case when
they are distinct, we now develop a
general
 method for 
representing the precise 
{\em sequence}
of arrivals of the first
few participants,
and show that these sequences have the potential to be very useful features in our prediction tasks.

For a comment thread $t$, let $t_i$ (for $i = 1, 2, ...$) denote
the identity of author of the $i^{\rm th}$ comment in the thread.
We now define the following encoding $\gamma(t)$ 
of comment thread $t$.
$\gamma(t)$ is a sequence of non-negative integers;
the $i^{\rm th}$ entry in the sequence, denoted $\gamma(t)_i$,
is equal to $0$ if $t_i$ is the author of the post that started the thread 
(returning to the thread in this case as the $i^{\rm th}$ commenter),
and otherwise $\gamma(t)_i$ is equal to the value of $j$
such that $t_i$ is the $j^{\rm th}$ distinct commenter to take part in $t$.
In what follows, we refer to $\gamma(t)_i$ as the 
{\em ID code} of commenter $t_i$.
We will also use the term {\em arrival pattern} to refer generically to any prefix of $\gamma(t)$ (including the full sequence).
Figure \ref{fig:patterns2} illustrates these concepts via two sample discussions, exemplifying that \focused threads should have arrival patterns in which some  back-and-forth between two participants is evident, whereas expansionary threads should have arrival patterns in which all  ID codes occur very few times, mostly just once.
\begin{figure}[h]
  \begin{center}
{\small
\newcommand{\comm}{\hspace{.1in}}
\begin{tabular}{l<{:} l| r<{:} l}
\multicolumn{2}{c|}{{\it \focused thread}} &  \multicolumn{2}{c}{{\it expansionary thread}} \\ \hline
\multicolumn{1}{l}{Mary:} & Anyone there & \multicolumn{1}{l}{James:} & we're engaged! \\
\comm Mary & ? & \comm Dina &  congrats! \\
\comm Don & me & \comm Fred & congrats! \\
\comm Pat & not me & \comm Mia & great!!! \\
\comm Don & v funny & \comm Moe & great! \\
\comm Pat & i know & \comm James & Thanks guys :) \\ 
\multicolumn{1}{r}{$\ldots$} & $\ldots$ & \multicolumn{1}{r}{$\ldots$}& $\ldots$ \\ \hline
\multicolumn{2}{l|}{{\it Length-2 arrival pattern: 0,1}} & \multicolumn{2}{l}{{\it Length-2 arrival pattern: 1,2}} \\
\multicolumn{2}{l|}{{\it Length-5 arrival pattern: 0,1,2,1,2}} & \multicolumn{2}{l}{{\it Length-5 arrival pattern: 1,2,3,4,0}}
\end{tabular}
}
    \caption{
 Example conversations  demonstrating our arrival-pattern coding scheme for the comment portion of threads.  
    \label{fig:patterns2}
     }
  \end{center}
  \vspace*{-0.25in}
\end{figure}

Can early (i.e., short) arrival patterns serve as useful features for our prediction tasks?  Before describing our full experiments (detailed in the next sections), it is useful to show some preliminary evidence of these patterns' potential utility.

\begin{figure}[h]
\fbox{{\includegraphics[width=3.2in,viewport=0 40 355 135,clip]{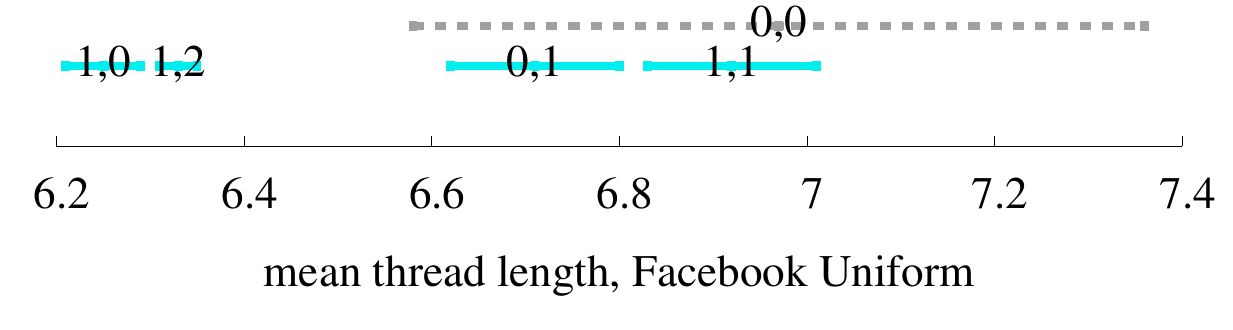}}} \\
\fbox{{\includegraphics[width=3.2in,viewport=0 40 355 135,clip]{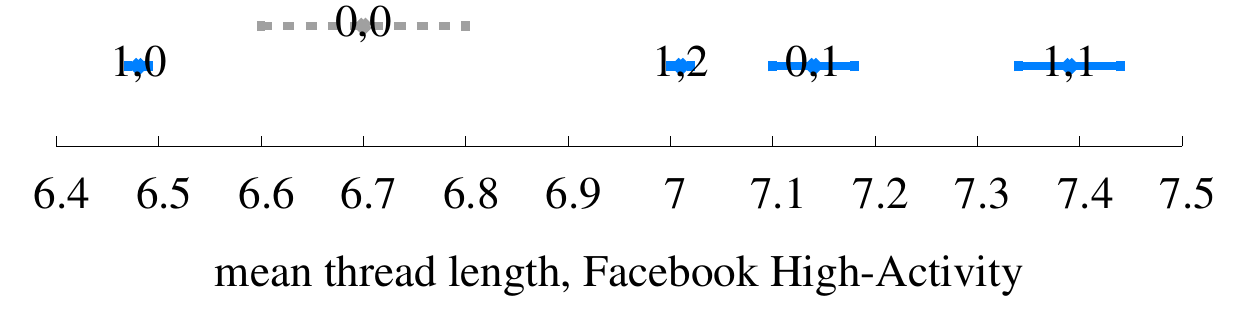}}}  \\
\fbox{{\includegraphics[width=3.2in,viewport=0 40 355 135,clip]{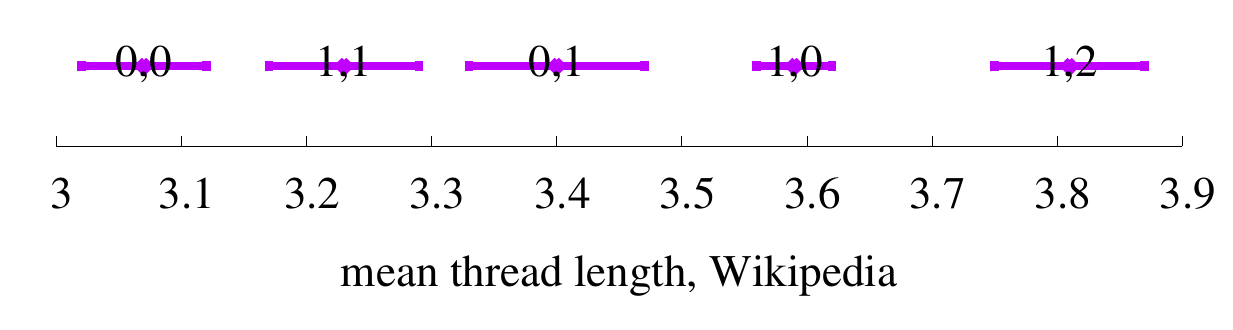}}} 
\caption{\label{fig:patterns-w}In all populations, the
  95\%-confidence-intervals for mean thread length for the five
  possible length-2 arrival patterns --- indicated as labels on the intervals
  --- are almost all disjoint.  Grey/dashed intervals indicate rare arrival patterns
  (at most 1\% of threads), so the long interval involved 
 in the single overlap (0,0
  in Facebook Uniform) is for a sparse situation.}
 \vspace*{-0.20in}
\end{figure}

First, we see whether different
(early)
 arrival  patterns tend to correspond to different thread lengths.
Figure \ref{fig:patterns-w}
shows,
for each of our three populations,
the (macro-averaged) length of 
threads
whose length-two prefixes correspond to each of the five possible length-two patterns;
the fact that the mean thread lengths fall in mostly disjoint confidence intervals indicates that the patterns do have predictive 
value.\footnote{ We note that the two most frequent arrival patterns in all three populations are (1,0) and (1,2), which is interesting because 
(1,0)  corresponds to the canonical turn-taking structure in a pairwise conversation,
while $(1,2)$ is the canonical sequence of successive new arrivals --- a 
further reflection of our \focused/expansionary dichotomy.
}

Second, we see whether different arrival patterns tend to correspond to differing re-entry probabilities, focusing on the chance that the user with ID code 1 (i.e., the first commenter who isn't the original poster) subsequently re-joins the thread by adding another comment.  
Table~\ref{tab:wiki_code_5} demonstrates that arrival patterns 
carry significant information about ID code 1's re-entry probability.  In all
the populations shown, it appears that 
guestbook-style patterns containing many
distinct ID codes tend to results in noticeably lower re-entry
probabilities.   For Facebook, we
also see a strong positive
correlation between the number of times ID code 1 appears in an
arrival pattern and the likelihood that ID code 1 will subsequently appear again.

\subsection{Timing effects} 
\label{sec:timing}

Our analysis thus far has considered the sequence of commenters
without any information about the speed at which they arrive in 
real time.  We now show some basic results establishing that
this type of temporal structure contains important information
about the length and re-entry properties of threads; in the next
section, we use this information as part of our prediction methods.

\begin{figure}[t]
  \begin{center}
    \includegraphics[width=0.4\textwidth]{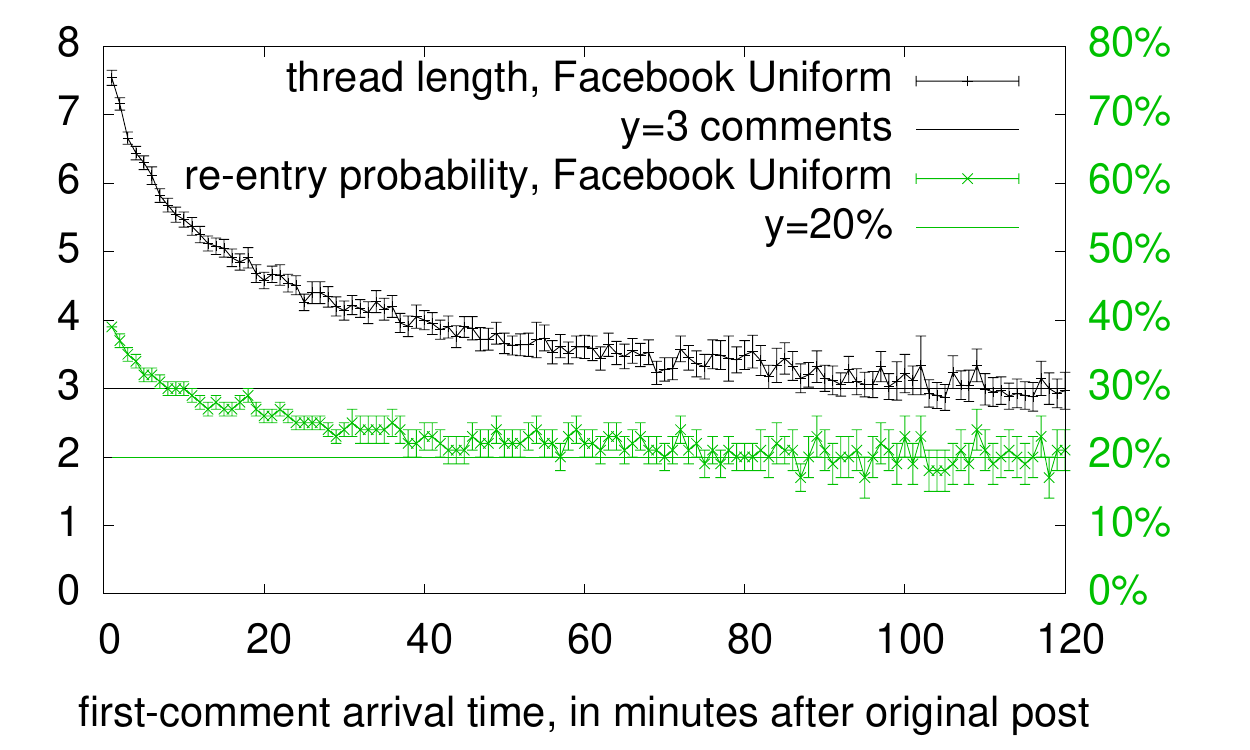}
    \caption{
Black (left axis, top curve): the longer it takes for the original post to attract its first
comment, the lower the expected thread length. Green (right axis,
bottom curve): in contrast, the probability that the first commenter
re-enters the thread is eventually independent of the first
comment's arrival time.  
The $y=3$ and $y=20\%$ lines are included to allow for visual comparison of the real data curves with theoretical curves in which the arrival lag has no effect.
    \label{fig:time-length}
    }
  \end{center}
  \vspace*{-0.02\textheight}
\end{figure}

In Figure \ref{fig:time-length}, we see (black
curve) that the longer it takes for the first comment to arrive on an
initial post, the shorter the thread, presumably because
``late'' first comments correspond to less overall activity around the post.
But note that timing isn't everything: beyond a certain
point, the probability that
the first commenter re-enters a thread (green curve) becomes 
approximately independent of the
first-comment arrival time lag.

\newcommand{\fnumwords}{num\_words} 
\newcommand{\fnumchars}{num\_chars} 
\newcommand{\textbase}{Text baseline}

\section{Predicting thread length}
\label{sec:predict-length}
We 
now
engage in 
the two main 
prediction tasks of this paper.
Recall that
the first task, which we describe in this section,  is to predict
thread length, as an indication of how much
interest a post will eventually generate,
given
the state of the thread at 
a certain early 
point.
 For example, we ask, given a 
thread
initiated by a Facebook user posting a status update to their friends that has already accumulated
5 comments, how well can we predict the final length of
the
thread?\footnote{ 
Naturally, for this task,  we use only features that can be derived
from the state of the thread when it had
5 comments.}
The 
results of the 
second task, to predict whether 
a user that has already participated in a thread will later re-enter that same thread, are described in 
 Section \ref{sec:predict-reentry}.

For thread-length prediction,
we formulate two concrete tasks on the Wikipedia and 
Facebook High-Activity datasets
(the results on the 
Facebook Uniform dataset are similar, but with larger error
bars).  
From the set of all posts made by the active Facebook users, we selected the subset of posts
that received at least 5 comments, and randomly reserved 50\% of them for
evaluation, using the other 50\% for feature exploration.  This gave us a test data
set of 1,996,624 posts.  Out of these, we chose a threshold of 8 comments to
create an approximately balanced binary
prediction problem: given the state of
the thread after five posts, will the thread eventually receive 
at least 8 comments?
(55.25\% of the posts are in the positive class.)
Similarly, for 
Wikipedia, we look at all 
talk-page posts that have received at
least two comments, and ask, will they receive a third
one?
 In this case, our data
is smaller, with only 44,732
items in the test set
(54.55\% of which are in the positive class).

{ 
\newcommand{\descr}[1]{#1} 
\newcommand{\ii}{i} 
\begin{table}[h]
\begin{center}
\begin{tabular}{|r| >{\raggedright}p{2.06in}|} \hline  
\multicolumn{2}{c}{LINKS} \tabularnewline \hline \hline
$edges\_prev[\ii]^*$ & Number of links from commenter to previous
commenters  \tabularnewline \hline
$mutual\_poster[\ii]^*$ & Number of links from commenter to users
linked to the original poster \tabularnewline \hline 
\multicolumn{2}{c}{ ARRIVAL PATTERNS} \\ \hline \hline
$id\_code[\ii]$ & commenter ID code as described in \S\ref{sec:patterns} \tabularnewline \hline
$uniq\_comm[\ii]$ & Unique commenters through comment $\ii$ \tabularnewline \hline
\multicolumn{2}{c}{TIME} \\ \hline \hline
$time[\ii]$ &  Time taken for the first $\ii$ comments to arrive \tabularnewline \hline
\multicolumn{2}{c}{TEXT REGRESSION FEATURES}   \\ \hline \hline
{\small $Orig\_post\_terms$} & ``comment'', ``agree'', etc.:
see \S\ref{sec:features}\tabularnewline \hline 
\multicolumn{2}{c}{MISC} \\ \hline \hline
$\fnumwords[i]$ & Number of words in comment $i$ \tabularnewline \hline
$\fnumchars[i]$ & Number of  characters in comment $i$ \tabularnewline \hline
$question[\ii]^*$ & Comment $\ii$ has a `?' \tabularnewline \hline
$exclaim[\ii]^*$ & Comment $\ii$ has a `!' \tabularnewline \hline
$likes[\ii]^*$ & Num likes on original post before comment $\ii$ is made \tabularnewline \hline
$comment\_likes[\ii]^*$ & Num likes on comments before comment $\ii$ \tabularnewline \hline
\end{tabular}
\end{center}
\caption{Features used in our prediction experiments.
 For each indexed feature, 
 we also build a comparable feature for the
 original post when it makes sense (the $id\_code$ for the original post 
is always 0 and
 so is omitted
but, for example,  the length of the original post in words or characters is meaningful).  Features marked with $^*$ were 
applied only
for Facebook data. 
 \label{tab:features}
}
\vspace*{-0.08in}
\end{table}
} 

\subsection{Features (used here and in \S\ref{sec:predict-reentry})}\label{sec:features}
The features we employed are 
summarized
in Table \ref{tab:features}.  
The first three sets are based on our  discussion above of links between
participants (\S\ref{sec:links}), arrival patterns
(\S\ref{sec:patterns}), and timing effects  (\S\ref{sec:timing}).  We
describe the other two sets now.

An important question is whether the textual features of the
original post are more or less effective for this task than the non-textual
features we have already described.  
To investigate this issue,
we elected to gather a small,
presumably general
set of such ``Original post terms'' features via text regression, 
which has
previously  been employed for blog comment-volume prediction \citep{Yano:ProcOfIcwsm:2010}. 
Specifically, we used
\hyphenation{Text-Regression}
J. M. 
White's TextRegression R package, which employs
linear regression with elastic-net regularization
\citep{Friedman:JournalOfStatisticalSoftware:2010}, run on a set of
posts disjoint from the training and test data used for classification.
50 terms were
selected for the Facebook data --- among them were ``comment''
and  ``anybody''
(positive coefficient for thread length), and ``re-post'' 
and URLs
(negative
coefficient). 
Among the 30 selected terms for Wikipedia were ``agree'' (positive) and ``thank'' (negative).

Also, preliminary pilot studies 
revealed a set of fairly intuitive miscellaneous
features, listed in the last section of Table \ref{tab:features},  that
are potentially
correlated with thread length.  For instance, one might expect that on
average, posts containing a question mark pose questions that prompt
comments as responses.

\subsection{Performance Results}
\label{sec:length-results}

Our testing methodology was: for a given set of features and
train/test set, create bagged
decision trees with 60 trees trained on independent samples of the
training data; then, apply the bagged decision trees on the disjoint
test set.   

Our main method was to use all the features described in
Table \ref{tab:features}.  We compared its performance against the
following two {\bf baselines}. The {\em positive-percentage bias
  baseline} chooses an item's label randomly with bias equal to the
percentage of test items in the positive class (55.52\% in the
Facebook case, 54.55\% in the Wikipedia case).  The {\em
  text-regression baseline} uses only  the $Orig\_post\_terms$ features
chosen via text regression as described in \S\ref{sec:features}.

The performance of our method versus the two baselines is shown in 
Table~\ref{tab:perf}. 
 Clearly, the combined use of participant-link,
arrival-pattern, timing, and other information yields the best results
for all five of our performance metrics.  The small set of
text-regression features extracted from the original post sometimes
did worse that the positive-percentage bias baseline.\footnote{This is
consonant with \citet{DeChoudhury:2009:MCI:1526709.1526754}, who
remark that ``textual analyses ... alone are not adequate to capture
conversational interestingness because
[they] do not consider the dialogue structure between users''.}

\begin{table}
\begin{center}
\begin{tabular}{|@{}l@{}|l|*{5}{r|}} \cline{3-7}
\multicolumn{2}{c|}{}                                      &ACC &   AUC & RMSE & APR & CXE \\ \hline \hline
\multirow{3}{*}{FB}   &Pos.-\% bias  baseline &   .552 & .500 & .497 & .550 & .992 \\
   &\textbase         & .537 & .529 & .503    & .568  & 1.01\\
   & All our features              & .\bf 672 & \bf .729 & \bf .457 & \bf .758 & \bf .872  \\ \hline \hline
\multirow{3}{*}{Wiki}   &Pos.-\% bias  baseline & .548 & .500 & .498 & .549 & .993 \\ 
   &\textbase &  .488 & .505 & .517 & .550 & 1.06\\
   & All our features                                                & \bf .595 & \bf .627 & \bf .486 & \bf .661 & \bf .958 \\ \hline
\end{tabular}
\end{center}
\vspace*{-0.010\textheight}
\caption {
Main thread-length prediction results.
Bold = best
performance per dataset,  under various metrics: ACC: accuracy 
(for FB active: after 5 comments, predicting whether the thread
achieves length $\geq$ 8; for Wiki: after 2 comments, predicting whether
an additional comment will occur). 
AUC:  area under the ROC curve. RMSE: root mean square error. APR: mean average
  precision. CXE: cross-entropy. 
\label{tab:perf}
}
\end{table}

\xhdr{Key Facebook Features}
To
better understand the
individual 
 factors contributing to the length of a comment thread,
we perform stepwise forward feature selection.  In iteration $j$ of this
algorithm, we create working feature set $F_j$ by finding the best single feature to add
to the set $F_{j-1}$ to maximize our objective function, 
area under the ROC curve (AUC).
Because it only selects a single feature at a time, this method
prevents us from adding more than a single copy of highly correlated features,
and the order that the features are installed gives us some insight into the
nature of these comment threads. 

\begin{figure}[t]
  \begin{center}
\includegraphics[width=0.3\textwidth]{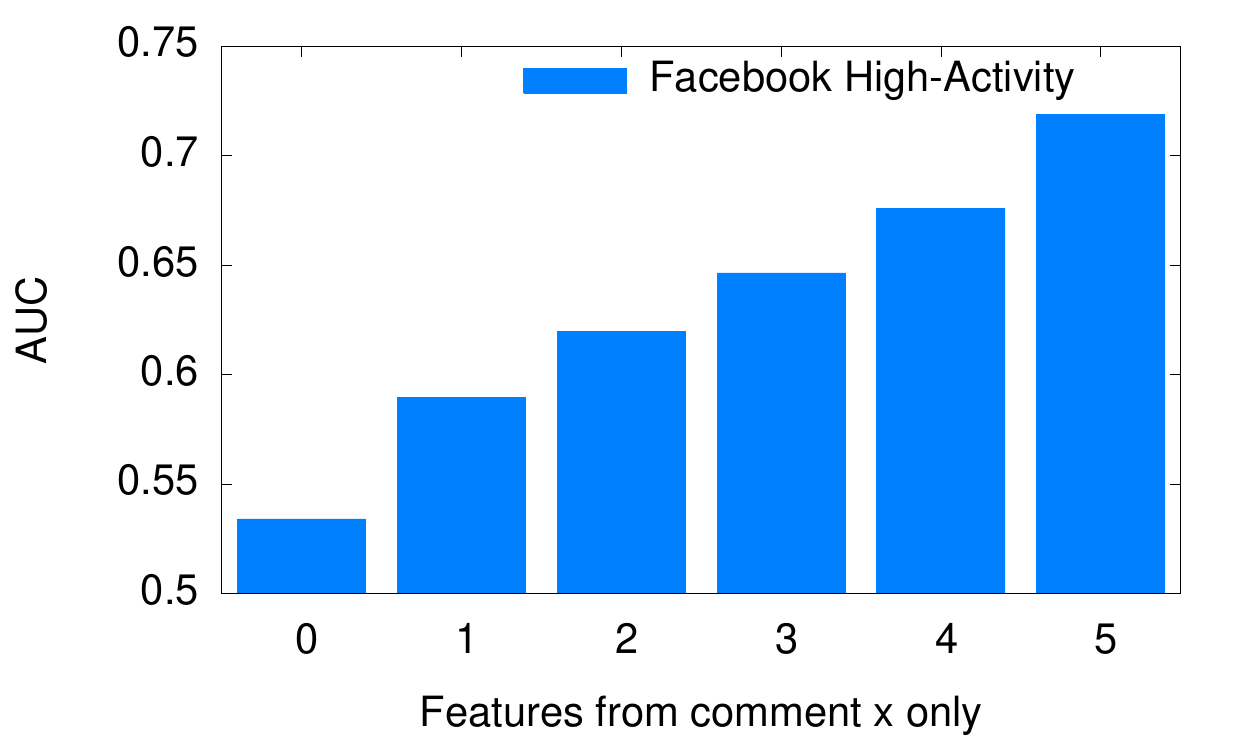} 
    \caption{Performance 
when predicting using only the features derived from a 
\emph{single}
    comment (or original post for $x = 0$).  The later the comment, the
    more informative. 
    \label{fig:bycom}
    }
  \end{center}
  \vspace*{-0.25in}
\end{figure}

Table~\ref{tab:featsel} shows the features
selected by this process for the Facebook dataset.
There are 
three
things worth noting from these results.  The first is that a
relatively small set of features contributes
almost
all of the predictive
value.  In particular,
the amount of time it takes for the first five comments to arrive 
(the TIME:$time[5]$ feature)
is highly
indicative of whether or not the thread will eventually reach 8
comments. Second, most of the key features come
from the fifth comment.  Thus, when predicting whether or not the thread will
continue, one should focus on the most recent activity of the thread.
We highlight this 
in
Figure~\ref{fig:bycom}, where we show the prediction performance when using only the
subsets of features derived from a single message in the thread, ranging from the
original post 
($x=0$) to the fifth comment.  
The third item of note is the fact that the link-based features 
do not have
much effect.  We believe this is because
they are low-recall, in the sense that we only showed in
\S\ref{sec:links} that they are useful when all the early commenters
in the thread are distinct.

Given
the strength of the time feature, it is interesting to ask
what would be the effect of its removal.
The  
combination of the other features is unable to make up for the loss of
temporal 
information:
removing that key feature, the AUC drops
from $0.729$ to $0.588$.  With or without that feature, and even if
we slice the data to predicting only for a fixed TIME:$time[5] \in [15m,20m)$, 
the relative ordering of the other features remains more or less unchanged.

\begin{table}
\begin{center}
\begin{tabular}{|l|c|} \hline
Feature added & AUC \\ \hline \hline
TIME:$time[5]$ & 0.6954 \\ \hline
+ARRIVAL PATTERN:$uniq\_comm[5]$ & 0.7053 \\ \hline
+MISC:$\fnumwords[5]$ & 0.7138 \\ \hline
+TIME:$time[3]$ & 0.7214 \\ \hline
+MISC:$question[5]$ & 0.7256 \\ \hline
+ARRIVAL PATTERN:$id\_code[5]$ & 0.7258 \\ \hline
+ARRIVAL PATTERN:$uniq\_comm[4]$ & 0.7260 \\ \hline
\end{tabular}
\end{center}
\caption {
Results of stepwise forward feature selection on Facebook. 
Each row represents performance for all features listed in that row and above.
 \label{tab:featsel}
}
\vspace*{-0.01\textheight}
\end{table}

\xhdr{Key Wikipedia Features}
In the case of Wikipedia, we see a somewhat similar ordering to the features.  
Again, the features 
regarding
the most recent comment 
(here, the second one)
are the most predictive of future
comments.  Here, however, we find that the
\emph{length} of the second comment is the
most important feature, followed by the time to the second comment,
and the
ID code of the second and first commenters.  Beyond these first four
features, the relatively small size of the dataset makes the predictive power
of other features unclear.

\section{Predicting thread re-entry}\label{sec:predict-reentry}

Here we examine 
our 
second, and novel,  prediction task: 
given the initial portion of a thread and the
identity of one of the commenters, how well can we predict whether
that commenter will contribute another comment later in the thread?
As noted 
earlier,
 the idea here is to determine whether to
keep a user notified of the progress on a thread after they have
commented on it already --- there are some threads for which the user
might want to actively return to the discussion. For space reasons, we
can only provide an overview of results here, omitting detailed
feature analysis.

For simplicity, we focus on the following two (related) questions.
Recall that we use ID code 0 for the original poster of a thread, and ID code 1
for the first commenter other than the original poster, assuming there
is such a commenter.  (a) Assuming that ID code 1 occurs in the
length-5 arrival-pattern prefix, does that user ever appear again?
(The value 5 was used in our thread-length prediction problem as
well.)  (b) The same, but for the first 9 comments.  
We use the same features as in
the previous task; see \S\ref{sec:features} and Table
\ref{tab:features}.

\begin{table}[h!]
\begin{center}
\begin{tabular}{|l|l|*{1}{r|}} \cline{3-3}
\multicolumn{2}{c|}{}                                      & AUC (x-val) \\ \hline
\multirow{3}{*}{FB (after 5 comments)}   &Pos.-\% bias  baseline &  .500  \\
   &\textbase      & .520 \\
   & Our features              & .\bf 808 \\ \hline 
\multirow{3}{*}{FB (after 9 comments)}   &Pos.-\% bias  baseline &  .500  \\
   &\textbase        & .525 \\
   & Our features              & .\bf 855 \\ \hline \hline
\multirow{3}{*}{Wiki (after 5 comments)}   &Pos.-\% bias  baseline & .500 \\ 
   &\textbase & .494 \\
   & Our features                                                & \bf .644 \\ \hline 
\end{tabular}
\end{center}
\vspace*{-0.010\textheight}
\caption {
Main thread-re-entry prediction cross-validation results. Bold marks the best performance per dataset. 
\label{tab:reentry-perf}
}
\end{table}

Using cross-validation, we find (Table \ref{tab:reentry-perf}) that the performance on the full feature set for Facebook is an
AUC of 0.855 for the 9 comment version, and 0.808 for the 5 comment
version
of the task.
Using the same feature selection methodology described above, we find that the most important
features are the identities of the individuals posting the comments
($id\_code[i]$),
and especially the identities of the most recent few commenters.
The time between the two most recent comments also plays an
important role, 
as the longer it takes, the slower the conversation is moving,
and the more likely it is to come to an end.

\def\mixp{\pi}
\def\seqp{{\bf p}}
\def\cmt{S}
\def\thfn{\theta}
\def\thgen{{\cal F}}
\def\thgenmix{{\cal F}^{(2)}}

\newtheorem{theorem}{Theorem}
\def\Prf{{\rm Pr}}
\def\ev{{\cal E}}
\def\evf{{\cal F}}

\newcommand{\Prb}[1]{
\Prf\left[{#1}\right]
}

\newcommand{\Prg}[2]{
\Prf\left[{#1}~|~{#2}\right]
}

\newcommand{\Exp}[1]{
E\left[{#1}\right]
}

\newcommand{\Expg}[2]{
E\left[{#1}~|~{#2}\right]
}

\newcommand{\prev}{re-entrant\xspace}
\newcommand{\prevs}{{\prev}s\xspace}

\section{Modeling Thread Re-Entry}
\label{sec:model}

Having gained some empirical understanding of thread re-entry, 
including relatively good performance at 
predicting it, we now seek to develop further theoretical
understanding of re-entry by formulating a set of
probabilistic generative models that produce arrival patterns
of a given fixed length.
We then study which of these models produce the qualitative
phenomena we observe in real threads --- particularly
bimodality in the number of distinct commenters.

The first class of models $\thgen$ we consider 
has the following basic structure 
for choosing who makes the $j^{\rm th}$ comment,
reminiscent of the Chinese Restaurant Process \citep{Aldous:1985a}.
With some fixed probability $p_j \geq 0$, 
we introduce a new participant;
with probability $1 - p_j$ we select, according to some underlying 
probabilistic rule, a participant who has already appeared in the
thread.  (We refer to such participants as {\em \prevs}).
The \prev selection rule is assumed to be a randomized 
algorithm $\thfn$ that takes a
thread prefix as input and produces the name of an existing participant
in the thread.
This is a very general definition;
depending on the choice of
the function $\thfn$, we can define arbitrary rules,
that, for example, pick
a \prev uniformly, or according to ``rich-get-richer''
principles that favor people who have commented more in the past
\cite{Kumar:2010:DC:1835804.1835875}, 
or according to recency principles so that an 
individual's selection probability decreases in the time since they 
last commented.
Each model $\Omega(k,\thfn,\seqp)$ in this class $\thgen$
is described by
a thread length $k$, selection rule $\thfn$, and sequence of probabilities 
$\seqp = p_1, p_2, p_3, \ldots, p_k$, $p_i \in (0,1]$.

\xhdr{A Negative Result} Although  $\thgen$ initially seems 
reasonable and covers a large space, 
it turns out to be a poor fit to reality,
because {\em none} of its members can yield 
the expansionary vs. \focused bimodality
that we found empirically in \S\ref{sec:dichotomy}.

\begin{theorem}
Let $\Omega(k,\thfn,\seqp)$ be an arbitrary model in the class $\thgen$,
and let $X$ be a random variable equal to 
the
number of distinct
participants in a length-$k$ thread $t$ generated by $\Omega(k,\thfn,\seqp)$
(counting the initial poster).
Then $X$ has a unimodal density function: there is a number $d^*$ such
that $\Prb{X = d}$ is monotonically increasing for $d \leq d^*$ and
monotonically decreasing for $d \geq d^*$.
\label{thm:unimodal}
\end{theorem}

We omit the proof due to lack of space, 
but it consists essentially of projecting the
arrival pattern onto a binary sequence that records only whether
each participant is a re-entrant or not.

\xhdr{Models Exhibiting Bimodality}
In view of this negative result, 
we seek an alternate class of models capable of generating arrival patterns
that exhibit bimodality in the number of distinct commenters. 

Arguably the simplest approach is to consider mixture models that have
bimodality ``built in":
We need only 
suppose that there are two distinct types of posts,
one which concentrates the number of distinct participants
on a small value, and the other which concentrates it on a large
value, and that threads are constructed by drawing one of the first type
with  fixed probability $\mixp > 0$ or one of the  second 
with probability $1 - \mixp$.

While this mixture principle is presumably an important reason why we see
bimodality in the real data, it is not the whole story.  Indeed, we
ran the following experiment to see whether
the same type of post can lead both to \focused and 
expansionary threads.  As it turns out, the CNN link that was most
shared among a large sample of Facebook users in the first quarter of 2012
was a report of Whitney Houston's death.  Although the set of threads
spawned just by shares of this link is small by the standards of Figure
\ref{fig:hm-random}, it is large in an absolute sense, and 
we observed in this controlled-content case
the same sort of  bimodality exhibited by threads overall:
sometimes, the news provoked  a series
of
 ``drive-by'' comments when it was shared by a user, and
other times, the same news prompted extended small-group
discussion. 

This finding motivates us to construct models of arrival patterns that
produce the expansion/\focus bimodality as a byproduct without assuming post type as its cause.
To do this, we posit a type of internal symmetry-breaking during thread
generation, taking inspiration from  the theory of nonlinear urn processes
\cite{arthur-lock-in}. In this new class of models, the probability that a new
participant enters at step $j$ depends on the identities of the participants
in the first $j-1$ steps. Intuitively, when there are many distinct
participants, the process should 
make re-entry less likely, thereby producing momentum in the 
expansionary direction;
when a few participants have each interacted multiple times,
the process should make it harder for new participants to break in,
thereby building up momentum in the conversational direction.

The class is parametrized by $\alpha \geq 1$ and $\beta \geq 0$.
For each participant $c$ 
already
in the  thread (including the original poster, 0),
and each length $j \leq k$, 
each such existing participant will have a 
{\em weight} $w_j(c)$ after step $j$ of the thread that controls
their probability of providing the next comment.  
The
fixed weight $\beta > 0$ 
controls the probability that
a new participant arrives in the next step.
We also impose the constraint that the same person never
appears twice in a row.\footnote{This is
essentially without loss of generality, since on the real
threads we can also build a comparable representation where we collapse out
consecutive occurrences of the same participant.}

Generating the arrival pattern $\gamma = \gamma_1 \cdots \gamma_k$
proceeds as follows.  
The first commenter will be labeled $1$ (since we do not have
the poster, labeled 0, provide the first comment too); so we
initialize by setting 
$\gamma_1 = 1, w_1(0) = w_1(1) = 1$, and following this initialization
we are positioned to determine the author of the second comment.
In general,
consider an arbitrary step  $j < k$, and let $c_j$ be the commenter in that step.  We proceed as follows.
\begin{enumerate}[(i)]
\item
{\em Choose commenter $j+1$}.
We choose a participant 
(different from $c_{j}$)
with probability proportional
to the weights.
Specifically: 
pre-existing
participant $c \neq c_j$ is chosen with probability 
$w_j(c) / (\beta + \sum_{c' \neq c_j} w_j(c'))$, and a 
new participant is introduced into the thread with probability 
$\beta / (\beta + \sum_{c' \neq c_j} w_j(c'))$.
We use $c_{j+1}$ to denote the participant chosen for step $j+1$.
\item\label{item:update}
{\em Update weights}.
If the participant $c_{j+1}$ in step $j+1$ is a \prev, we define $w_{j+1}(c_{j+1}) = \alpha w_j(c_{j+1})$,
and leave all other weights unchanged.
If 
instead  $c_{j+1}$ is
new, we define $w_{j+1}(c_{j+1}) = 1$ and for all other 
pre-existing
participants 
$c \neq c_{j+1}$ we reduce their weights by setting
$w_{j+1}(c) = w_j(c) / \alpha$.
\end{enumerate}
The key point is the weight update rule in part (\ref{item:update}). A new arrival
suppresses the weight of all existing participants, making it less
likely they will comment again and paving the way for further new arrivals.
On the other hand, when an existing participant provides the next comment,
their increase in weight makes it more likely they will return,
thereby promoting back-and-forth interaction.

\begin{figure}[t]
  \begin{center}
    \includegraphics[width=0.30\textwidth]{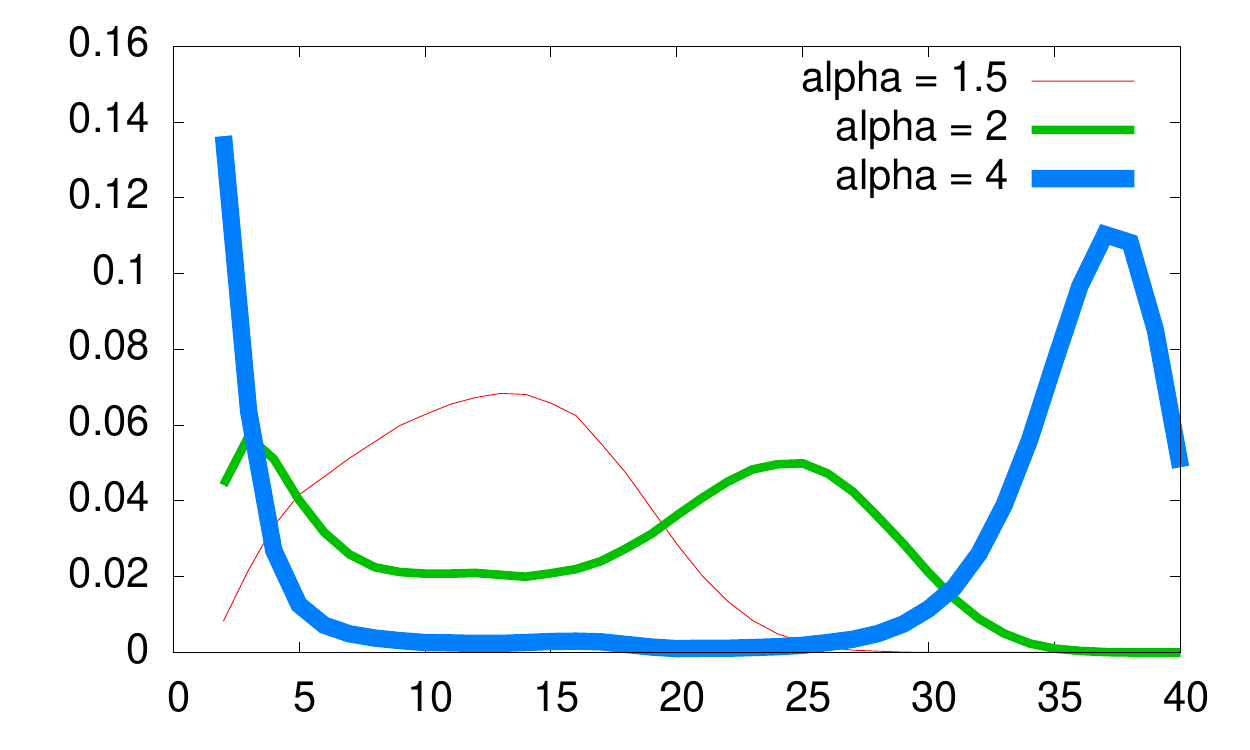}
    \caption{Density function of distinct participants in threads produced
       by 
our proposed family of processes
with $\alpha = 1.5$, $2$, and $4$ 
       ($\beta = 1$ and $k = 40$).
    \label{fig:bimodal}
    }
  \end{center}
\vspace*{-.03\textheight}
\end{figure}

We show via simulation that bimodality emerges naturally in this model.  
To paraphrase Langston Hughes, the number of distinct participants
can dry up like a raisin in the sun, or it can explode.
Figure \ref{fig:bimodal} shows the empirical
density function obtained through simulation
for the number of distinct participants under multiple settings of the
model parameters: we fix the length $k = 40$ and $\beta = 1$, and
then we simulate the process with $\alpha = 1.5$, $2$, and $4$.
As we see there, bimodality emerges as $\alpha$ increases, which accords
with intuition --- larger values of $\alpha$ are more aggressive in 
amplifying both the focused and expansionary effects, and hence
serve to bifurcate the process into its two modes more strongly.

The model appears to be quite challenging to analyze rigorously, 
and it is an interesting open question to prove that it produces
bimodality, as well as to characterize
the transition from unimodality
to bimodality as we increase $\alpha$.
The model shares some properties with nonlinear urn processes
\cite{arthur-lock-in}, but also has ingredients that
lie beyond what is usually needed for the analysis of such processes.

\section{Related Work}
\label{sec:relwork}

To our knowledge, there has not been prior consideration in the
literature of the overall problem of algorithmic conversation curation
--- an emerging key component in enhancing user experience in current
forms of on-line social interaction.  This problem involves many
issues, including those investigated in this paper: {\bf (a)} determining
which posts are interesting enough to bring to a user's attention; {\bf
  (b)}
among discussions a user already has knowledge of, choosing which the
user should continue to be updated about; and, indirectly, {\bf (c)}
understanding the structure of discussions, both to aid in the two
issues just described and potentially for implications in
user-interface design. Of course, there has been much valuable work on
the first and last issue individually, which we now describe. (Our
attention to {\bf (b)} appears to be novel.)
 On {\bf (a)}, we point out DeChoudhury et al.'s research
\citep{DeChoudhury:2009:MCI:1526709.1526754} on the interestingness of
Youtube comment threads, as measured by interestingness of topic and
participants (not length), and Shmueli et al.'s work
\citep{DeChoudhury:2009:MCI:1526709.1526754} on predicting which
stories a particular user is most likely to comment on. Prior work on
comment-volume prediction \citep{Tsagkias:2009:PVC:1645953.1646225,
Yano:ProcOfIcwsm:2010, Guerini:ProceedingsOfIcwsm:2011,
Wang:ProceedingsOfKdd:2012,artzipredicting} is of course also quite
relevant. How fast a piece of information spreads or diffuses
\citep{Kwak:2010:TSN:1772690.1772751,Lerman:ProceedingsOfIcwsm:2010,Bakshy:ProceedingsOfWsdm:2011,Romero:ProceedingsOfThe20ThInternationalConferenceOn:2011,artzipredicting}
is another important aspect of interestingness.  Quality of posts or comments, as
determined by ratings, is potentially also relevant;  see for example \citet{Siersdorfer:ProceedingsOfWww:2010}.
On {\bf (c)}, there is intriguing work
\citep{Gomez+Kaltenbrunner+Lopez:2008a,Kumar:2010:DC:1835804.1835875}
on structural characterizations of discussions when viewed as trees
(not an approach we have taken in this paper, and arguably less
natural as a model for discussions on sites like Facebook that
have post-and-comment-interfaces).  Different
perspectives are taken by researchers looking at characterizations of
agreement and/or sentiment among comments
\cite{Mishne+Glance:2006a,Gilbert:HawaiiInternationalConferenceOnSystemSciences:2009,Park:ProceedingsOfCscw:2011,Chmiel:PlosOne:2011},
and by sociological analyses of turn-taking in 
group conversations \cite{Gibson01062010}.

\begin{figure}[t]
  \begin{center}
  \includegraphics[width=0.35\textwidth]{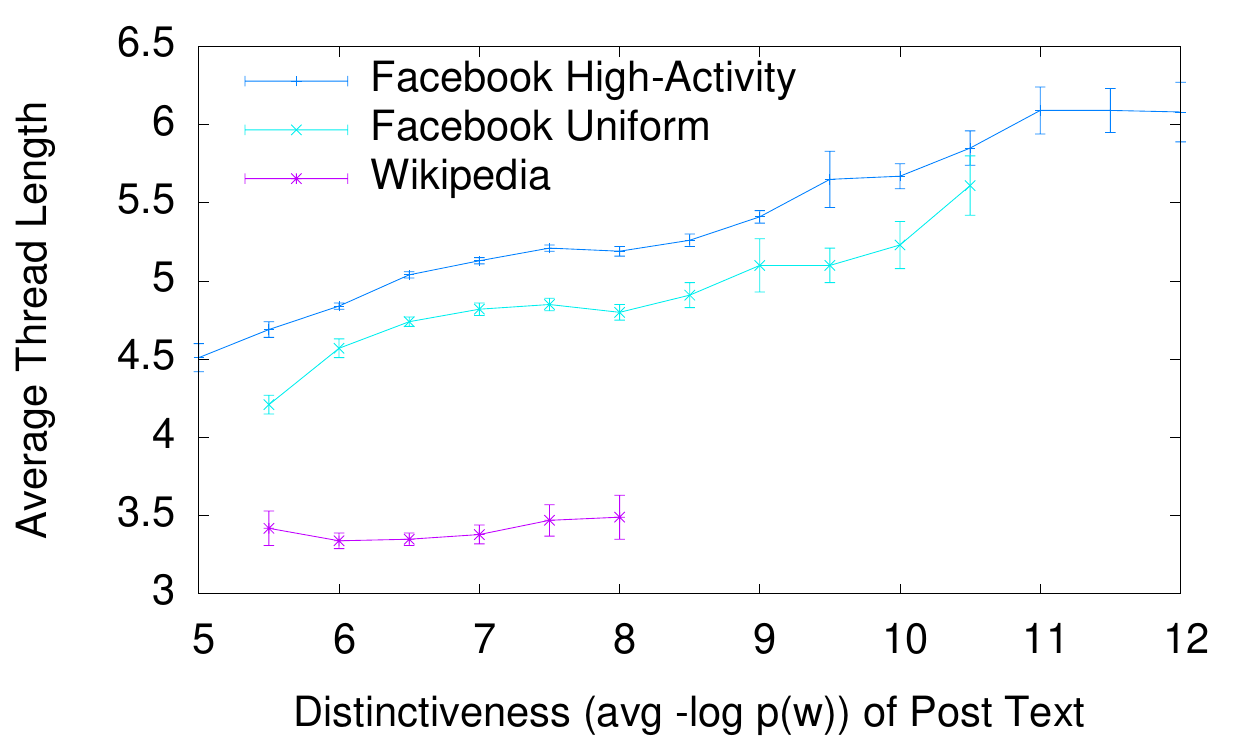}
    \caption{Considering only 8+-word  posts that generated responses, for Facebook, the more distinctive
      the text of the original post, the more comments it garners, but
      for 
      Wikipedia, which  is more task-oriented, there is no such effect.
    \label{fig:prob-go}
    }
  \end{center}
  \vspace*{-0.25in}
\end{figure}

\section{Future Work: Distinctiveness}
\label{sec:typical}

We see many exciting future directions to pursue. 
Here, we briefly highlight two preliminary explorations into 
{\em distinctiveness features} that we
believe hold promise, although we have not yet 
identified a way of applying them in their current form
to improve prediction performance.
The main idea is that the  likelihood of a post's
text or early commenters should be informative.

\yhdr{Distinctiveness of Text}

A basic property of a piece of text is its {\em likelihood} ---
whether its word choices look typical when compared to a reference
collection, or whether its word choices are less likely
and hence more {\em distinctive}.
In recent work, measures of distinctiveness were shown to help in recognizing 
movie quotes that were deemed ``memorable'' in the sense of
cultural penetration 
\cite{Danescu-Niculescu-Mizil:2012:YMH:2390524.2390647}.
In our case, the question is the following:
When a post contains unusual text, what should this lead us to estimate about
the length of the resulting comment thread?
There are intuitive arguments in both directions: 
some low-probability posts might generate
discussion because they are provocative and unexpected, 
but others might simply be hard to understand and thus be mainly ignored.

We built a unigram language model from 3.5 million Facebook posts
by authors whose posts weren't in our main dataset;
for each word $w$, the model provides a probability $p(w)$.
We define a post's {\em distinctiveness} to be
the average over its  tokens $w$ of 
$\log (1/p(w))$; lower distinctiveness means a more likely post.
Figure \ref{fig:prob-go} shows the macro-averaged
post length as a function of text distinctiveness, considering {\em only}
posts containing at least 8 words\footnote{At 8 words and beyond,
 post distinctiveness becomes empirically almost independent of post length,
 disentangling the two features.} and  that received at least one comment in the case
of Facebook or at least two for Wikipedia.
For these particular subsets, our two Facebook
populations exhibit a
clear positive effect of the distinctiveness of the text, whereas for
Wikipedia there seems to be no effect at all (which perhaps stems from the task-oriented nature of Wikipedian discussions).  
We note, however, that the effects
become less clear if we include posts that turned out to
generate no comments (or at most one on Wikipedia).

\begin{figure}[t]
  \begin{center}
    \includegraphics[width=0.4\textwidth]{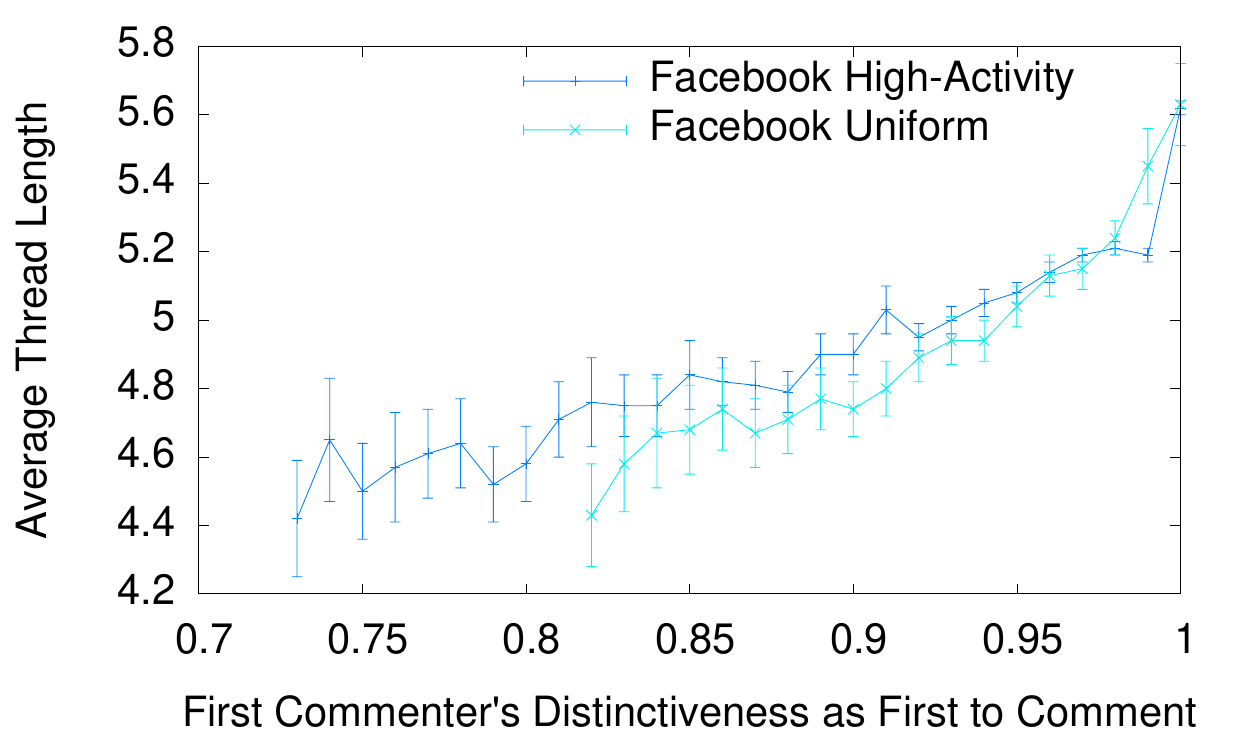}
    \caption{In Facebook, the more distinctive the first commenter is
      (in terms of not often being first to respond to the original
      poster), the longer the thread. Wikipedia is not depicted due to
      sparseness, but the overall trend is the opposite.
    \label{fig:group-plot}
    }
  \end{center}
  \vspace*{-0.25in}
\end{figure}

\yhdr{Distinctiveness of First Commenter}
For a user $u$ who posts regularly, a set of frequent commenters
on $u$'s threads often emerges --- the people who generally
weigh in when $u$ says something.
Thus, shifting from the likelihood of words to the likelihood of users,
it makes sense to ask
about the effect on thread length of the first commenter's distinctiveness ---
the extent to which this commenter is usually or rarely 
first in one of $u$'s threads.
Again, there are intuitive arguments each way: 
if the first commenter $v$ is someone who's often a first commenter on
$u$'s posts,  then $v$
is presumably familiar to both $u$ and 
the audience for $u$'s comments, which could make it easier
for the thread to grow; but it may also be socially easier
to let $v$'s comment pass by without much 
activity.

In Figure \ref{fig:group-plot} we show for Facebook the expected thread length
as a function of the fraction of times the first commenter was  {\em
  not} the
first to respond to the original poster's
posts\footnote{For the uniform population, the plot only consider users 
with at least ten posts,
although different threshold values do not greatly affect the
resulting trends.}.
We see a clear upward trend:
when someone you rarely hear from first is in fact the
first to comment on your post,
on average it foreshadows a longer thread, perhaps because this
indicates that the post has greater reach.

We note that 
Wikipedia appears to exhibit the opposite behavior.  
We do not depict the Wikipedia results
due to sparseness
of recurring first commenters, 
but when restricting to users
with at least 10 posts and binning the distinctiveness values, 
we see a significant decreasing trend.

\section{Conclusions}

Motivated by the growing role of automated mechanisms to manage
users' interactions with on-line discussions, we have identified
and studied two key problems in the curation of such discussions.
The first of these, length prediction, is related to earlier studies
of comment volume on blog and news sites, but it acquires additional 
complexity in our context due to the heterogeneity we find in long threads,
which can either be focused on a few participants or expand to reach many.
The second problem, re-entry prediction, has to our knowledge not
been formulated previously; it is a crucial issue in applications that
must decide when to notify users about updates to discussions in which
they have participated.

We see these two problems as helping to define the contours of the 
problem of conversational curation more broadly, and as such the results
here suggest a range of further open questions.
Among these are a deeper understanding of the features that can help
predict the trajectory of an on-line discussion from its early stages,
and the integration of these techniques into systems that deliver
discussion-oriented content to users in on-line applications.

\smallskip

 \small
\noindent {\bf Acknowledgments} We thank Tom Lento and the anonymous reviewers
for their helpful comments.
Supported in part by NSF grant
 IIS-0910664.
This work was done while the last author was at Cornell University.
\bibliographystyle{abbrv-acm-shrink}
\providecommand*{\bibfont}{\raggedright}

\end{document}